\documentclass[12pt]{iopart}

\usepackage{color,graphicx,iopams}

\usepackage[english]{babel}

\usepackage[unicode=true,colorlinks=true]{hyperref}

\hypersetup{linkcolor=blue,citecolor=blue,urlcolor=blue}

\definecolor{mygrey}{gray}{0.35}
\definecolor{myblue}{rgb}{0.2,0.2,0.8}
\definecolor{myzard}{cmyk}{0,0,0.05,0}
\definecolor{mywhite}{rgb}{1,1,1}
\definecolor{mywhite}{rgb}{1,1,1}
\definecolor{myred}{rgb}{1,0.,0.3}

\begin{document}

\title{Effective many-body Hamiltonians of qubit-photon bound states}

\author{T. Shi$^{1,2}$, Y.-H Wu$^{2,3}$, A. Gonz\'{a}lez-Tudela$^{2}$ and J. I. Cirac$^{2}$}

\address{$^{1}$ Institute of Theoretical Physics, Chinese Academy of Sciences, Beijing 100190, China, \\ $^{2}$  Max-Planck-Institut f\"ur Quantenoptik, Hans-Kopfermann-Stra\ss{}e 1, 85748 Garching, Germany,  \\ $^{3}$ School of Physics, Huazhong University of Science and Technology, Wuhan 430074, China}

\ead{yinghaiwu88@hust.edu.cn}
\ead{alejandro.gonzalez-tudela@mpq.mpg.de}

\begin{abstract}
Quantum emitters (QEs) coupled to structured baths can localize multiple photons around them and form qubit-photon bound states. In the Markovian or weak coupling regime, the interaction of QEs through these single-photon bound states is known to lead to effective many-body QE Hamiltonians with tuneable but yet perturbative interactions. In this work we study the emergence of such models in the non-Markovian or strong coupling regime in different excitation subspaces. The effective models for the non-Markovian regime with up to three excitations are characterized using analytical methods, uncovering the existence of doublons or triplon states. Furthermore, we provide numerical results for systems with multiple excitations and demonstrate the emergence of polariton models with optically tuneable interactions, whose many-body ground state exhibits a superfluid-Mott insulator transition.
\end{abstract}

\maketitle

\newpage

\section{Introduction}

Quantum systems coupled to a common environment experience interactions mediated by the bath excitations~\cite{cohenbook92a}. In quantum electrodynamics, this is the basic mechanism behind the forces between electrons, atoms, or molecules. Those interactions can be tuned if one controls the coupling to the bath, which opens up exciting possibilities in quantum information. One of the most prominent examples in this context is that of two-level quantum emitters (QEs) coupled to the free-space electromagnetic field, resulting in the well known dipole-dipole interactions between QEs~\cite{lehmberg70a,lehmberg70b}. Unfortunately, these interactions are generally accompanied by spontaneous emission, which limits their applications. The latter can be avoided if the density of modes at the transition frequency vanishes, since the spontaneous decay rate is proportional to that quantity. This occurs, for instance, if one embeds QEs in a cavity such that QEs are far-off resonance from the cavity modes~\cite{raimond01a,ritsch13a}.

Another way of cancelling spontaneous emission while obtaining exchange interactions among QEs is by endowing the bath with a periodic structure, which strongly influences the density of states~\cite{bykov75a}. In fact, band gaps where the density of states vanishes can emerge, so that spontaneous emission in that bath can be prevented, but still interactions between the emitters can be mediated by virtual processes via the common bath~\cite{john90a,kurizki90a}. Experiments with atoms, quantum dots, or superconductors interacting with structured bath has renewed the interest in investigating these phenomena~\cite{thompson13a,goban13a,lodahl15a,hood16a,liu17a,mirhosseini18a,sundaresan18a}. Other experimental scenarios involving cold atoms in optical lattices with state-dependent potentials are amenable to the same description, and thus the appearance of analogous phenomena have been predicted~\cite{devega08a,navarretebenlloch11a} and have recently been observed~\cite{krinner17a}.

The theoretical description of the interactions mediated by a bath is relatively simple in the so-called Markovian limit~\cite{lehmberg70a,lehmberg70b,agarwal77a}. There, it is possible to derive a master equation for the quantum emitters only, where the bath degrees of freedom are traced out. This effective description contains both Hamiltonian and dissipative Lindblad terms. The latter vanishes if the QEs transition frequency lies in the band gap, whereas the first one describes the interactions between the QEs mediated by the bath, where one of them is excited when another one is de-excited. The emergence of dipole-dipole interactions in this scenario, which opens the door to investigate spins models with long range interactions~\cite{shahmoon13a,douglas15a,gonzaleztudela15c}, can be attributed to the existence of a single-photon bound states~\cite{john90a}. An intuitive picture~\cite{douglas15a} is that the single-photon bound state acts as an off-resonant cavity mode that mediates interaction between the QEs. The strength and functional form of these interactions depend on the QE-bath coupling strength, detuning as well as the band-dispersion relation. Although, these interactions can be made relatively strong to overcome other dissipative mechanism, their predicted strength is ultimately limited by the Markovian conditions under which these effective description has been derived.

In this work, we study the effective many-body Hamiltonians emerging when QEs couple to structured baths beyond the Markovian limit, and investigate to which point the dipole-dipole description survives in this regime. Furthermore, we study the consequences on the effective QE interactions of the emergence of multi-photon bound states, which were recently predicted in the single QE regime~\cite{shi09a,sanchezburillo14a,shi16a,calajo16a,sundaresan18a}, but which impact in the multi QE situation has not yet been fully considered.  Our analysis allows us to uncover qualitatively different interaction Hamiltonians, in which multi-photon bound states (doublons/triplons) hop from QE to the other, and analytically characterize them up to three excitations. With more excitations, we numerically characterize the emergent polariton models using density matrix renormalization group (DMRG), and show that their interactions can be controlled optically through the QE-bath interactions, allowing us to probe the quantum phase transition between superfluid and Mott insulator.

This manuscript is organized as follows. In Sec.~\ref{sec:model}, we explain the model that will be used throughout the paper and review the results in the Markovian limit to have them as reference for the next Sections. In Sec.~\ref{sec:1excitation}, we study the single excitation subspace, deriving an effective Hamiltonian to describe the dipole-dipole interaction for two and many QEs. In Secs.~\ref{sec:GF}-\ref{sec:2EDL}, we study the two-excitation subspace for both the two and many QE regimes. Since the phenomenology in this regime differs significantly from what is expected in a Markovian description, we use Sec.~\ref{sec:GF} to explain the analytical tools we use to characterize it, and describe qualitatively the main features that emerge in this subspace, namely, the scattering of two polariton modes and the hopping of doublon states. This emergent dynamics will be discussed in detail in Secs.~\ref{sec:2EGS} and~\ref{sec:2EDL}, respectively.  In Sec.~\ref{sec:manyexcitation}, we analytically characterize the three excitation subspace, and then go to the many-excitations regime to numerically explore a superfluid to Mott-insulator phase transition appearing in the ground state of the systems. Finally, we summarize our results and conclude in Sec.~\ref{sec:conclu}.

\section{Setup and Markovian limit\label{sec:model}}

\begin{figure}[tbp]
\centering
\includegraphics[width=0.7\linewidth]{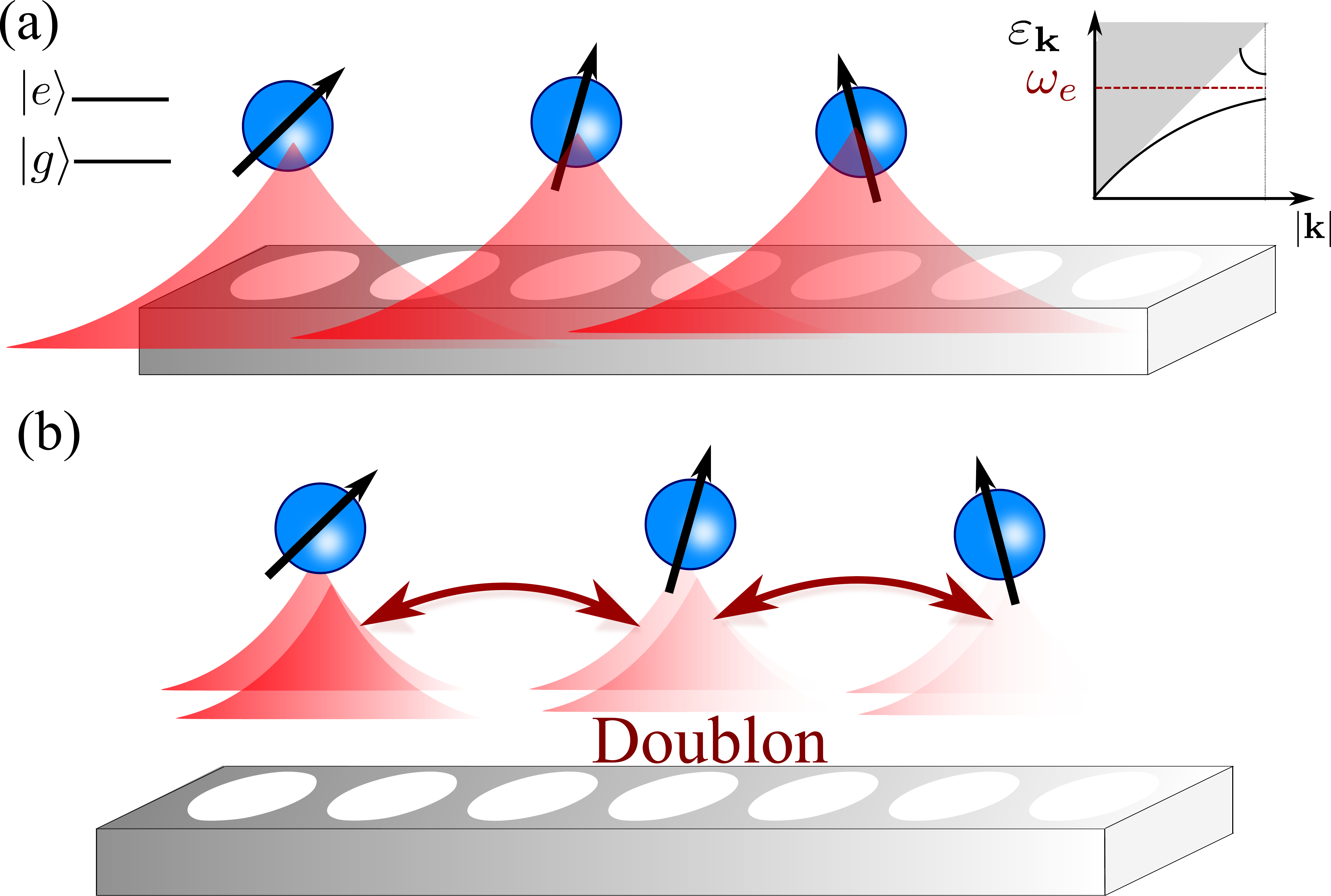}
\caption{(a) Scheme of the model: Several QEs are coupled to a structured bosonic bath with energy dispersion $\varepsilon_\mathbf{k}$. Though, we schematically plotted a 1D bath, the results can be extended to higher dimensions. (b) Pictorial representation of the doublon states hopping between QEs.}
\label{fig1}
\end{figure}

\subsection{Model}

As shown schematically in Fig.~\ref{fig1}(a), we consider $N_{b}>1$ QEs with two energy levels $\{|g\rangle_{j},|e\rangle_{j}\}$ ($j=1,\dots,N_{b}$) coupled to a structured bath. The total Hamiltonian of the system reads (with $\hbar=1$):
\begin{equation}
H = H_{A} + \sum_{\mathbf{k}} \varepsilon_{\mathbf{k}} a_{\mathbf{k}}^{\dagger
} a_{\mathbf{k}} + \frac{\Omega}{\sqrt{N}} \sum_{\mathbf{k},j} (e^{-i\mathbf{k}\cdot\mathbf{n}_{j}} a_{\mathbf{k}}^{\dagger} \sigma_{ge}^{j} + \rm{H.c.}),
\label{eq:Ham}
\end{equation}
where $H_{A}=\omega_{e}\sum_{j=1}^{N_{b}}\sigma_{ee}^{j}$ describes an array of $N_{b}$ two-level QEs with transition frequency $\omega_{e}$. Periodic boundary conditions (PBCs) for the bath are used, so we can label its modes in terms of the quasi-momentum $\mathbf{k}$. Here, $a_{\mathbf{k}}$ ($a_{\mathbf{k}}^{\dagger}$) is the annihilation (creation) operator of the bath mode with quasi-momentum $\mathbf{k}$, $\varepsilon_{\mathbf{k}}$ is the energy dispersion relation, $N$ is the total number of bath modes, and $\Omega $ is the QE-bath coupling strength. We will assume that the bath has a single band of width $W$, although the results can be easily extended to other situations. To obtain analytical results, we will go into the continuum limit, where $N\rightarrow\infty$ so that $\mathbf{k}$ becomes a continuous vector, and we can replace sums by integrals. Finally, $\sigma_{\alpha\beta}^{j}=|\alpha\rangle_{j}\langle\beta|$ are the spin operators for the $j$-th QE and $\mathbf{n}_{j}$ its corresponding vector position.

In optical and microwave implementations, we require that $\Omega\ll\omega_{e},\rm{min}\varepsilon_{\mathbf{k}}$, so the counter-rotating terms for the QE-bath coupling can be neglected. For convenience, we will work in the rotating frame at the frequency $\rm{min}\varepsilon_{\mathbf{k}}$, which amounts to taking
\begin{equation}
H_{A}=\Delta \sum_{j=1}^{N_{b}}\sigma _{ee}^{j}
\end{equation}
where $\Delta=\omega_{e}-\rm{min}\;\varepsilon_{\mathbf{k}}$. A crucial feature of the Hamiltonian in Eq.~(\ref{eq:Ham}) is that the number of excitations, defined as
\begin{equation}
N_{\rm{exc}}=\sum_{\mathbf{k}}a_{\mathbf{k}}^{\dagger}a_{\mathbf{k}}+\sum_{j}\sigma_{ee}^{j}\,,
\end{equation}
is conserved. This allows us to derive effective models separately in the subspaces with different numbers of excitations. In this work, we will concentrate on the few-body scattering and bound-state behaviors in the subspaces with $N_{\rm{exc}}=1,2,3$, and the quantum phase transitions in the ground states of the subspaces with many excitations.

Though most of the expressions we derive are valid for an arbitrary energy dispersion $\varepsilon_{\mathbf{k}}$ (see Appendices), in the main text we focus on the results for a 1D tight-binding model, where the energy dispersion reads:
\begin{equation}
\varepsilon_{k}=2J-2J\cos k,
\label{1Ddispersion}
\end{equation}
with $J$ being the hopping strength and $k=0,2\pi/N,\ldots,2\pi(N-1)/N$.

\subsection{QEs as hard-core bosons}

For the calculations performed in this paper, it is convenient to describe QEs using hard-core bosons. In this representation, we replace $\sigma_{ge}^{j}\rightarrow b_{j}$, where $b_{j}$ is an annihilation operator fulfilling bosonic commutation relations, and we restrict the Hilbert space to the states with $(b_{j}^{\dagger})^{2}=0$. In practical terms, this can be done by writing:
\begin{equation}
H_{A} = \Delta\sum_{j=1}^{N_{b}} b_{j}^{\dagger}b_{j} + \frac{U}{2}\sum_{j=1}^{N_{b}} b_{j}^{\dagger}b_{j}^{\dagger}b_{j}b_{j}
\end{equation}
and taking the $U\rightarrow\infty$ limit, which forbids double occupation of the $b_j$ modes.

In the case of many QEs, we assume that they are equally spaced, which allows us to work in Fourier space by defining $b_{j}=\sum_{\mathbf{p}}b_{\mathbf{p}}e^{i\mathbf{p\cdot }\mathbf{n}_{j}}$. The quadratic part of the Hamiltonian is $H_{0}=\sum_{\mathbf{p}}H_{\mathbf{p}}$ with
\begin{equation}
H_{\mathbf{p}}=\sum_{\mathbf{K}}\varepsilon_{\mathbf{k}}a_{\mathbf{k}}^{\dagger}a_{\mathbf{k}}+\Delta b_{\mathbf{p}}^{\dagger}b_{\mathbf{p}}+\frac{\Omega}{\sqrt{z}}\sum_{\mathbf{K}}(a_{\mathbf{k}}^{\dagger}b_{\mathbf{p}}+b_{\mathbf{p}}^{\dagger }a_{\mathbf{k}}) 
\label{Hp}
\end{equation}
that commute with each other (i.e., $[H_{\mathbf{p}},H_{\mathbf{p}^{\prime}}]=0$), where the photon momentum $\mathbf{k=p+K}$ is given by the QE quasi-momentum $\mathbf{p}$ and the reciprocal momentum $\mathbf{K}$ of the sublattice, and $z=N/N_{b}$ is the number of bath modes per unit cell. The hard-core interaction part $H_{\rm{hc}}$ becomes
\begin{equation}
H_{\rm{hc}} = \frac{U}{2N_{b}} \sum_{\mathbf{p}_{1},\mathbf{p}_{2},\mathbf{q}} b_{\frac{\mathbf{q}}{2}+\mathbf{p}_{1}}^{\dagger} b_{\frac{\mathbf{q}}{2}-\mathbf{p}_{1}}^{\dagger} b_{\frac{\mathbf{q}}{2}-\mathbf{p}_{2}} b_{\frac{\mathbf{q}}{2}+\mathbf{p}_{2}}.
\end{equation}

\subsection{Markovian limit}

Let us here remind the results obtained in the Markovian limit when the QEs transition frequency lies within the band-gap, which means that $\Delta<0$, and $|\Delta|,W\gg\Omega$. In that limit, one can eliminate the bath degrees of freedom using standard quantum optics techniques and obtain an effective Hamiltonian for the QEs \cite{lehmberg70a,lehmberg70b}. For two QEs with relative position $\mathbf{d}=\mathbf{n}_{2}-\mathbf{n}_{1}$, one obtains:
\begin{equation}
H_{2}^{\rm{M}} = H_{A} + V_{\rm{dd}}(\mathbf{d})(\sigma _{1}^{+}\sigma_{2}^{-}+\sigma_{2}^{+}\sigma_{1}^{-}),
\label{Markov2}
\end{equation}
where $V_{\rm{dd}}(\mathbf{d})$ is the dipole-dipole interaction strength, which in the limit $N\rightarrow\infty$, reads:
\begin{equation}
V_{\rm{dd}}(\mathbf{d})=\Omega^{2}\int \frac{d\mathbf{k}}{(2\pi)^{D}}\frac{{e^{i\mathbf{k}\cdot\mathbf{d}}}}{\Delta-\varepsilon_{\mathbf{k}}}
\end{equation}
for a general $D$-dimensional bath and $\sigma_{j}^{+}=(\sigma_{j}^{-})^{\dagger}\equiv\sigma_{eg}^{j}=(\sigma_{ge}^{j})^{\dagger}$. For many QEs, one obtains
\begin{equation}
H_{N_{b}}^{\rm{M}} = H_{A} + \sum_{i,j}V_{\rm{dd}}(\mathbf{d}_{i}-\mathbf{d}_{j})(\sigma_{i}^{+}\sigma_{j}^{-}+\sigma_{j}^{+}\sigma_{i}^{-}).
\label{MarkovMany}
\end{equation}
Note that the dipole-dipole interaction for many QEs is the same as for two QEs under the Markov approximation. For the 1D tight-binding model, one obtains \cite{shahmoon13a}
\begin{equation}
V_{\rm{dd}}=-\frac{\Omega^{2}}{|\Delta|}e^{-d/\xi} 
\label{Vdd}
\end{equation}
with the decay length $\xi=\ln^{-1}(|\Delta|/J)$. We will use these effective Hamiltonians as a baseline to compare with the results of the next sections. In particular, we will see what is the regime of validity, and how it has to be modified outside that regime. To do that, we will first study analytically up the three-excitation manifold, and finally perform DMRG calculations for the case with many excitations.

\section{Single Excitation \label{sec:1excitation}}

In this section we study the physics of the single excitation subspace. This regime has been extensively studied in the literature (see, for instance, Ref.~\cite{calajo16a} and references therein). Here, we will review results that are relevant for the other sections, and also derive simple formulas for the emergent effective models.

We divide this section in two parts: in Section~\ref{TQE} we study the situation when only two QEs are coupled to the bath, deriving an effective exchange interaction Hamiltonian valid in a particular region of the $(\Delta/J,\Omega/J)$ parameter regime that we will define. In Section~\ref{subsec:QEarray1exc}, we study the situation with many QEs and derive an effective hopping model in the lowest band, which can be defined in all parameter regimes. In both regimes, when $|\Delta|\gg\Omega$ we recover the effective spin models predicted for the Markovian limit in Eqs.~\ref{Markov2}-\ref{MarkovMany}. However, in the strong-coupling limit, $\Omega\gg{J},\Delta$, an effective spin model can still be derived to characterize the hopping of the strong hybridized QE and photon, i.e., a polariton, where the effective hopping strengths are dramatically enhanced compared to those in the Markovian regime.

\subsection{Two QEs\label{TQE}}


The single-excitation eigenstates for the system with two QEs can be generally written as $|\Psi_{1\lambda}\rangle=\beta_{\lambda }^{\dagger}|0\rangle$ with
\begin{equation}
\beta_{\lambda}^{\dagger} = u_{1,\lambda}b_{1}^{\dagger }+u_{2,\lambda}b_{2}^{\dagger}+\sum_{\mathbf{k}}f_{\lambda}(\mathbf{k})a_{\mathbf{k}}^{\dagger}.
\end{equation}
The coefficients $u_{1,\lambda}$, $u_{2,\lambda}$, and $f_{\lambda}(\mathbf{k})$ for all the eigenstates (including bound states and scattering states) can be obtained by solving the Schr\"{o}dinger equation $H|\Psi_{1\lambda}\rangle=E_{1\lambda}|\Psi_{1\lambda}\rangle$ (see~\ref{app:1ES}). The probability weights of symmetric and anti-symmetric modes $b_{\pm}^{\dagger}=(b_{1}^{\dagger}{\pm}b_{2}^{\dagger})/\sqrt{2}$ in the eigenstate $|\Psi_{1\lambda}\rangle$ are $Z_{1\lambda}^{\sigma=\pm}=|\langle{0}|b_{\sigma}\beta_{\lambda}^{\dagger}|{0}\rangle|^{2}$.

In general, the two lower eigenstates, $|\Psi_{1\lambda=\pm}\rangle=\beta_{\pm}^{\dagger}|0\rangle$, represent the symmetric and anti-symmetric superpositions of two local single-excitation bound states around the QEs. However, as we will show below (and already derived in Ref.~\cite{calajo16a}), only the symmetric bound state survives in certain parameter regimes. In the regime where both bound states exist, the Hamiltonian can be projected into the subspace spanned by these two bound states, which gives rise to the following effective model for the low energy part of the spectrum:
\begin{equation}
H_{\rm{eff}}^{(1)} = E_{1+}\beta_{+}^{\dagger}\beta_{+} + E_{1-}\beta_{-}^{\dagger}\beta_{-}. 
\label{Hs}
\end{equation}

A basis transformation converts this Hamiltonian into a hopping model
\begin{equation}
H_{\rm{eff}}^{(1)} = E_{0}\sum_{j=1,2} \widetilde{\beta}_{j}^{\dagger}\widetilde{\beta}_{j} + t_{\rm{eff}}(\widetilde{\beta}_{1}^{\dagger}\widetilde{\beta}_{2}+\rm{H.c.}),
\label{Hst}
\end{equation}
for the two localized Wannier modes $\widetilde{\beta}_{1,2}=(\beta_{+}\pm\beta_{-})/\sqrt{2}$, where $E_{0}=(E_{1+}+E_{1-})/2$ is the effective chemical potential and $t_{\rm{eff}}=(E_{1+}-E_{1-})/2$ is the effective hopping strength. 

If the distance $d$ between two neighboring QEs is sufficiently large, the two nearly degenerate ground states with $E_{1+}{\sim}E_{1-}$ describe the single-excitation bound states localized around two individual QEs. These two bound states begin to hybridize with each other as $d $ decreases. When the distance $d$ is much smaller than the localization length of the bound states, the strong mixing of the bound states induces the energy level splitting $2t_{\rm{eff}}$ around $E_{0}$. If the splitting is large enough, the anti-symmetric bound state energy merges in the continuum and the state is not bound anymore. For example, for the 1D bath with the tight-binding dispersion relation, the symmetric bound state always exists, but the anti-symmetric bound state can only be found in the regime $\Delta<\Omega^{2}d/2$~\cite{calajo16a}. Obviously, the effective hopping model of Eq.~\ref{Hst} can not be defined when only the symmetric bound state exists. In this paper, we mainly focus on the regime where both bound states exist, such that the low-energy physics of the single excitation is described by the effective Hamiltonian (\ref{Hst}), where the relevant parameter is the hopping strength $t_{\rm{eff}}$.

\begin{figure}[tbp]
\centering
\includegraphics[width=0.7\linewidth]{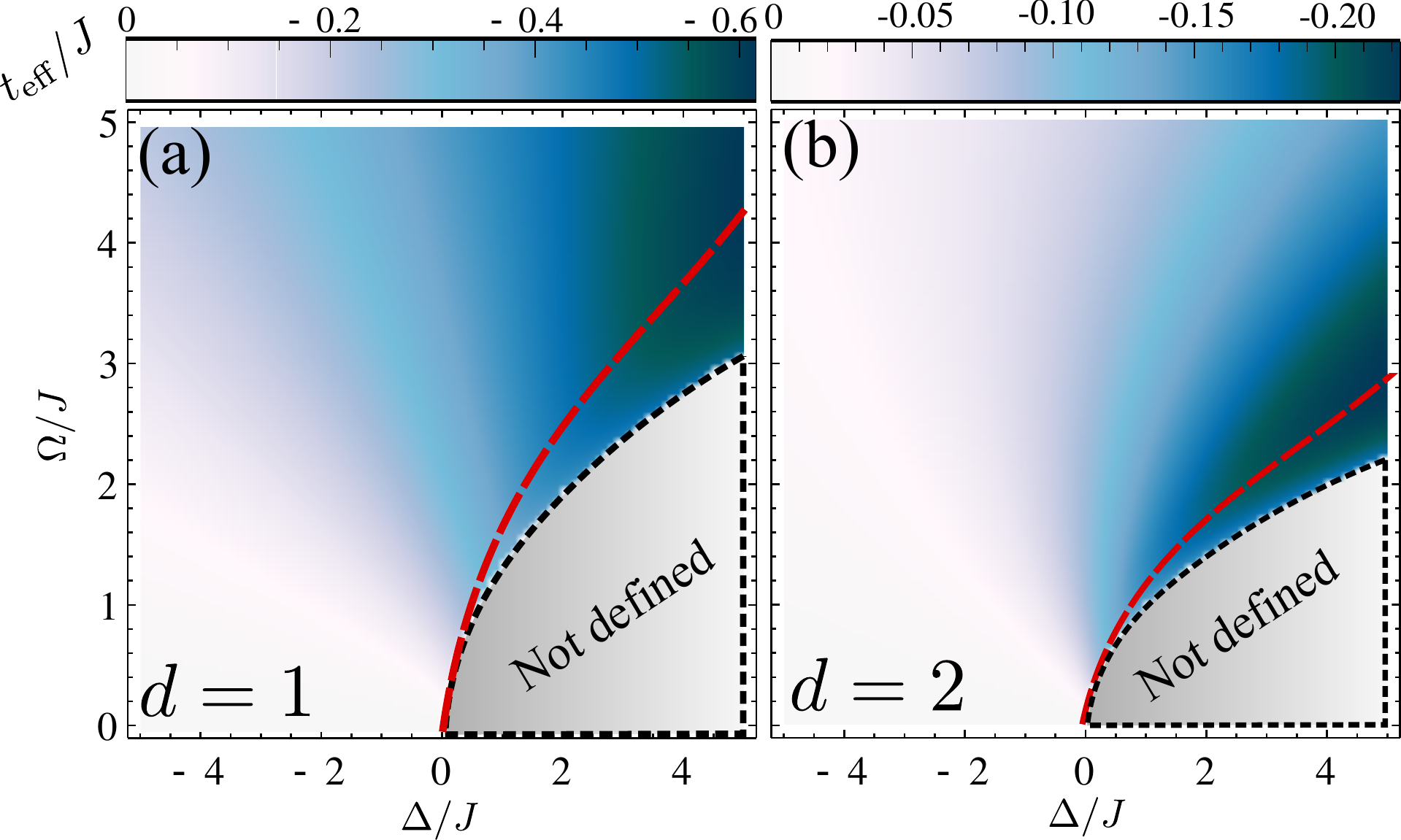}
\caption{The effective hopping strength, $t_{\rm{eff}}=(E_{1+}-E_{1-})/2$, in the $\Delta-\Omega$ plane for the 1D tight binding dispersion relation. $J$ is chosen as the unit and (a) $d=1$ (b) $d=2$.}
\label{twoatomse}
\end{figure}

As shown in Fig.~\ref{twoatomse}, in the Markovian $|\Delta |\gg\Omega $ and strong coupling $\Omega\gg J,\Delta$ regimes, a parameter regime that we denote as the {arc} region, the photon is strongly localized around QEs. Thus, the effective chemical potential $E_{0}\sim E_{1B}$ tends to the energy $E_{1B}$ of the bound state around a single QE. The effective hopping strength when $t_{\rm{eff}}{\ll}E_{0}$ can be approximated by $t_{\rm{eff}}{\sim}-Z_{1B}\Omega^{2}e^{-d/\xi}/\sqrt{E_{1B}(E_{1B}-4J)}$ (see~\ref{app:1ES}), where the single-particle weight and the decay length are
\begin{eqnarray}
Z_{1B} &=& \left[ 1+\Omega^{2}\int\frac{dk}{2\pi}\frac{1}{(E_{1B}-\varepsilon_{k})^{2}} \right]^{-1}, \nonumber \\
\xi &=& \left\{ \ln \frac{1}{2} \left[2-\frac{E_{1B}}{J}+\frac{1}{J}\sqrt{E_{1B}(E_{1B}-4J)}\right] \right\}^{-1}.
\end{eqnarray}

In the Markovian limit, the effective hopping strength $t_{\rm{eff}}{\sim}-J_{\rm{eff}}|\Delta/J|^{1-d}$ decays exponentially with the distance $d$, where $J_{\rm{eff}}=J\Omega^{2}/\Delta^{2}$, such that it reproduces the result of Eq.~(\ref{Vdd}). In the strong coupling limit, the effective hopping strength $t_{\rm{eff}}=-J_{\rm{eff}}(\Omega/J)^{1-d}$ also decays exponentially, however, $J_{\rm{eff}}=J/2$, which means that the coupling strengths are significantly enhanced due to the strong hybridization between QE excitation and photon.

In Fig.~\ref{twoatomse}, the effective hopping strength $t_{\rm{eff}}$ is shown in the $\Delta-\Omega$ plane for two QEs with $d=1,2$ coupled to the 1D tight-binding bath. For $\Delta<0$, $|t_{\rm{eff}}|$ increases monotonically and saturates to $J/2$ in the strong coupling limit. For $\Delta>0$, $|t_{\rm{eff}}|$ increases to the maximal value at $\Omega_{\rm{max}}$ slightly above the boundary $\Omega=\sqrt{2\Delta/d}$, and decreases to $J/2$ in the strong coupling limit $\Omega \gg J,\Delta$. The Rabi frequency $\Omega_{\rm{max}}$ maximizing $|t_{\rm{eff}}|$ is highlighted by the dashed red lines in Fig.~\ref{twoatomse}.

\subsection{QE array\label{subsec:QEarray1exc}}

Let us now consider the situation where we have now many QEs coupled to every $z$ bath lattice sites. By imposing PBCs for the QE array, the excitations will have $z+1$ bands for the QE propagation, although here we focus on the lowest energy band.  Using the intuition from the previous Section (see also~\cite{calajo16a}), we expect that the bath mediates QE interactions giving rise to a polariton propagating in the QE lattice. Let us now explain how to characterize this emergent behaviour.

\begin{figure}[tbp]
\centering
\includegraphics[width=0.7\linewidth]{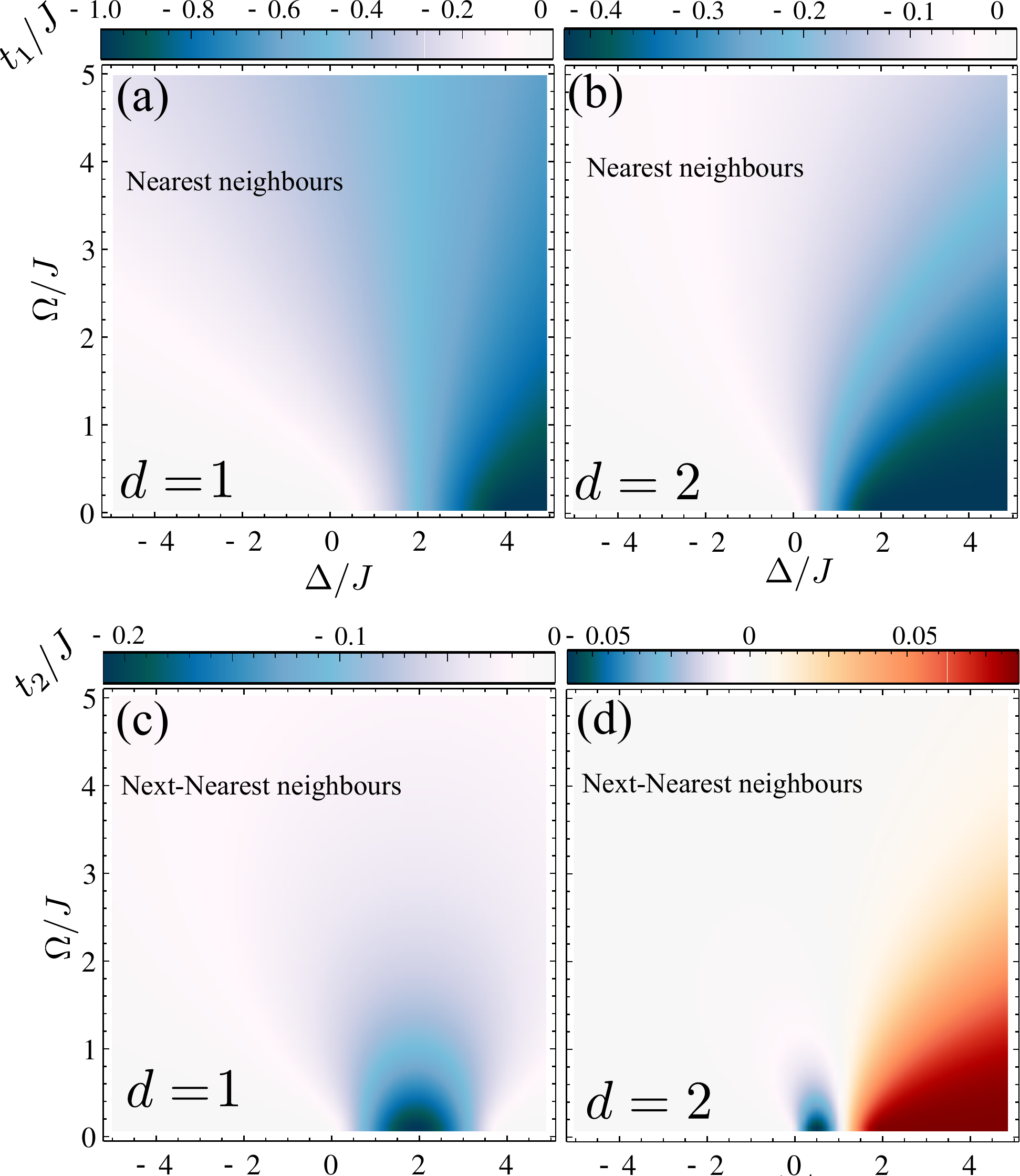}
\caption{The effective NN and NNN hopping strengths for the 1D tight binding dispersion relation of bath: (a)-(b) The NN hopping strength in the $\Delta-\Omega$ plane for $d=1$ and $d=2$, respectively. (c)-(d) The NNN hopping strength in the $\Delta-\Omega$ plane for $d=1$ and $d=2$, respectively.}
\label{manyatomse}
\end{figure}

Since we are restricting in this Section to the single excitation subspace, the hard-core interaction Hamiltonian plays no role, and therefore the single-particle modes with different quasi-momenta $\mathbf{p}$ in the first Brillouin zone are decoupled. Thus, the QE-bath Hamiltonian is quadratic and can be diagonalized, $H_{0}=\sum_{\mathbf{p}\lambda}E_{1\lambda}(\mathbf{p})\beta_{\mathbf{p}\lambda}^{\dagger}\beta_{\mathbf{p}\lambda}$, by the annihilation (creation) operator $\beta_{\mathbf{p}\lambda}$ ($\beta_{\mathbf{p}\lambda}^{\dagger}$) of the single polariton with momentum $\mathbf{p}$ and dispersion relation $E_{1\lambda}(\mathbf{p})$ in the $\lambda$-th band, where $\lambda\in[0,1,2,\cdots,z]$ labels the different energy bands. In the polariton mode $\beta_{\mathbf{p}\lambda}^{\dagger}|0\rangle$, the probability weight of QE mode of quasimomenta $\mathbf{p}$ being in the excited state is given by:
\begin{equation}
Z_{1\lambda }(\mathbf{p})=|{\langle}0|b_{\mathbf{p}}\beta_{\mathbf{p}\lambda}^{\dagger}|0\rangle|^{2}.
\end{equation}

The dynamics in the lowest band (i.e., $\lambda =0$) is given by:
\begin{equation}
H_{\rm{eff}}^{(1)} = PH_{0}P = \sum_{\mathbf{p}}E_{1}(\mathbf{p}) \beta_{\mathbf{p}}^{\dagger} \beta_{\mathbf{p}},
\end{equation}
where $P$ is the projector into the states $\beta_{\mathbf{p}}^{\dagger}|0\rangle{\equiv}\beta_{\mathbf{p}\lambda=0}^{\dagger}|0\rangle$ in the lowest band $\lambda=0$, and  $E_{1}(\mathbf{p}){\equiv}E_{1\lambda=0}(\mathbf{p})$ its energy dispersion relation. The Wannier modes can be obtained via Fourier transform as $\widetilde{\beta}_{j}=\sum_{\mathbf{p}}\beta_{\mathbf{p}}e^{i\mathbf{p}\cdot\mathbf{n}_{j}}/\sqrt{N_{b}}$, which describes the local single-excitation bound state around the $j$-th QE. In terms of these local Wannier modes, the effective Hamiltonian in the coordinate space can be written as
\begin{equation}
H_{\rm{eff}}^{(1)} = \sum_{jj^{\prime}}t_{j-j^{\prime}}\widetilde{\beta}_{j}^{\dagger}\widetilde{\beta}_{j^{\prime}}
\end{equation}
with hopping strengths
\begin{equation}
t_{j-j^{\prime }}=\frac{1}{N_{b}}\sum_{\mathbf{p}}E_{1}(\mathbf{p})e^{i\mathbf{p}\cdot(\mathbf{n}_{j}-\mathbf{n}_{j^{\prime }})}.
\end{equation}
In the arc region, the dominant hopping constant $t_{j}\sim-t_{1}\delta_{j1}$ can be deduced from the first order degenerate perturbation theory, where $t_{1}=J(1-Z_{1B})$. In the Markovian and large Rabi coupling limit, $t_{1}=J\Omega^{2}/\Delta^{2}$ and $J/2$ agrees with $J_{\rm{eff}}$ in the two-QE case.

In the large $d$ limit, the vanishing $t_{j}{\rightarrow}0$ indicates that the local Wannier modes reduce to the single-excitation bound states localized around different individual QEs. As the lattice spacing decreases, $t_{j}$ becomes finite and the local Wannier modes hybridize with each other to form the dispersive polariton band. Here, we note that since PBC for the QE array is applied, the translational symmetry ensures that the number of states in the lowest band is the same as the QE number due to the Bloch theorem. However, if the QE array has finite size and we have open boundary condition, like it occurs for two QEs coupled to the bath, some polariton modes may vanish in certain parameter regimes due to the boundary effect, as we showed in previous section.

In Fig.~\ref{manyatomse}, the nearest neighbor (NN) and next-nearest neighbor (NNN) hopping strengths $t_{l=1,2}$ in the $\Delta-\Omega$ plane are shown for the 1D tight-binding dispersion relation. In the arc region of the $\Delta-\Omega$ plane, the Wannier mode in the lowest band is the single excitation bound state strongly localized around each individual QE, and the overlap of two Wannier modes exponentially decays with the localization length $\xi{\leq}1$. As a result, the NN hopping strength $t_{1}{\sim}t_{\rm{eff}}$ can be determined by the hopping strength in the two-QE case, and the long range hopping strength $t_{l>1}{\sim}0$. In the Markovian limit $|\Delta|\gg\Omega$, the single particle weight $Z_{1\lambda=0}(\mathbf{p}){\equiv}Z_{1}(\mathbf{p}){\sim}1$ shows that the polariton state in the lowest band is mostly composed of QE excitations, and the band is only slightly deformed from a completely flat one. In the strong coupling limit $\Omega{\gg}J,\Delta$, the reduced $Z_{1}(\mathbf{p}){\sim}1/2$ exhibits the strong hybridization of QE excitation and bath photon modes, which gives rise to the finite width$\sim 2J$ of the polariton band.

Non-Markovian effects emerge in the intermediate regime, where the long range hopping strengths $t_{l>1}$ in general do not vanish, as shown in Figs.~\ref{manyatomse}c and~\ref{manyatomse}d. In the small $\Omega/J$ limit, the hopping strengths saturate to $t_{1}/J=-0.424$ and $t_{2}/J=0.085$ for $d=2$ when the detuning $\Delta /J>2$. This saturation can be understood using the single-excitation band structure in the limit of $\Omega{\rightarrow}0$. If $0<\Delta<\Delta_{c}{\equiv}\varepsilon_{k=\pi/d}$ and $\Omega/J$ is small, the lowest band states are mostly composed of photons in the bath and the lowest band has a width $\sim\Delta$. If $\Delta>\Delta_{c}$, the width of the lowest band saturates to $\Delta_{c}$ at small $\Omega$, and $t_{l}{\rightarrow}2(-1)^{l}\sin(\pi/d)/[\pi{d}(l^{2}-d^{-2})]$.

Summing up, we have derived and characterized the effective hopping model emerging in the single-excitation subspace when many QE are coupled periodically to the bath modes. In particular, we have shown it reduces to an effective tight-binding model in the arc region of the $\Delta-\Omega$ plane. In the Markovian regime $|\Delta|\gg\Omega$ [strong coupling limit $\Omega{\gg}J,\Delta$], the effective model describes the propagation of the bare QE excitation [the hybridized QE-photon polariton excitation with the single particle weight $1/2$]. In the latter, the effective NN hopping is significantly enhanced compared to that in the Markovian regime.

\section{General features of two excitation subspace \label{sec:GF}}

The dynamics with more than one QE excitation can be characterized by the effective spin model of Eq.~\ref{MarkovMany} only in the Markovian limit. To the best of our knowledge, the use of that effective model beyond the Markovian regime is not justified, and how to characterize the dynamics in the whole $\Delta$-$\Omega$ plane is not clear yet.  In this section, we pedagogically review the tools on how to treat both the two and many QE situation in the two-excitation sector, and explain qualitatively the emergent features. In particular, we will see that two different types of excitations appear in the two-excitation subspace, namely, the scattering states of two single polaritons and the \emph{doublon} states formed by two bound polaritons, as we show schematically in Fig.~\ref{fig1}(b). Then, we discuss them in detail in Sections~\ref{sec:2EGS} and~\ref{sec:2EDL}, respectively.

\begin{figure}[tbp]
\centering
\includegraphics[width=0.8\linewidth]{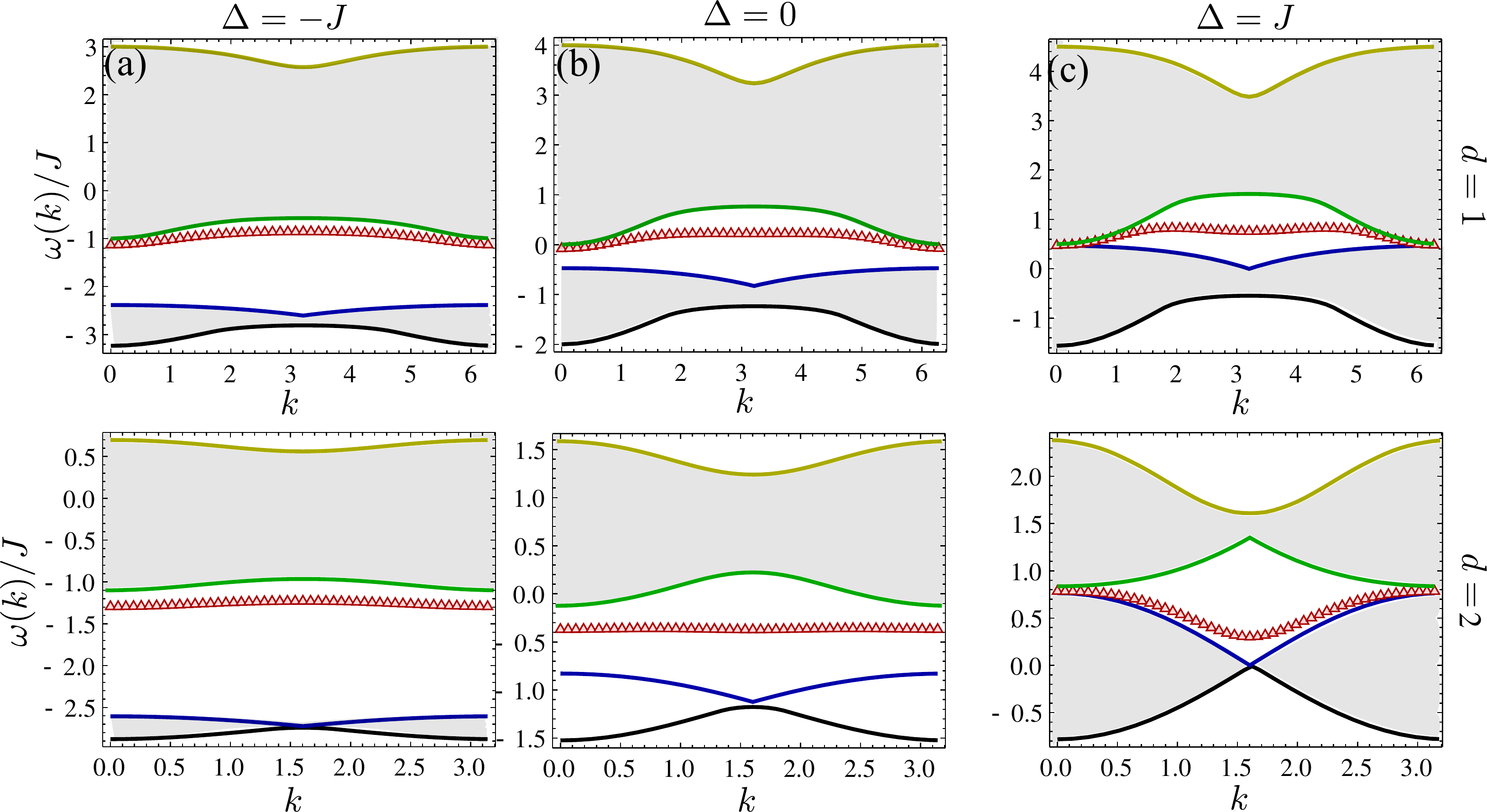}
\caption{For the 1D tight-binding dispersion relation of photons, the first three bands including two scattering bands and one doublon band in the middle. (a)-(c) for $d=1$: $\Delta/J=-1$, $\Omega/J=1$ (a); $\Delta/J=0$, $\Omega/J=1$ (b); $\Delta/J=1$, $\Omega/J=1$ (c); (d)-(f) for $d=2$: $\Delta/J=-1$, $\Omega/J=1$ (d); $\Delta/J=0$, $\Omega/J=1$ (e); $\Delta/J=1$, $\Omega/J=1$ (f).}
\label{bandstructure}
\end{figure}

\subsection{Two QEs}

When only two QEs are coupled to the bath, the general two-excitation eigenstate has the form
\begin{eqnarray}
|\Psi_{2\lambda}\rangle &=& \sqrt{Z_{2\lambda}} b_{1}^{\dagger}b_{2}^{\dagger} |0\rangle + \sum_{\mathbf{k},j=1,2} \varphi_{j\lambda}(\mathbf{k}) a_{\mathbf{k}}^{\dagger} b_{j}^{\dagger}|0\rangle + \sum_{\mathbf{kk}^{\prime}} \varphi_{2\lambda}(\mathbf{k},\mathbf{k}^{\prime})a_{\mathbf{k}}^{\dagger}a_{\mathbf{k}^{\prime}}^{\dagger}|0\rangle. 
\label{p2}
\end{eqnarray}
The eigenvalue $E_{2\lambda}$ of $|\Psi_{2\lambda}\rangle$, the probability weight $Z_{2\lambda}$, and the wavefunctions $\varphi_{1\lambda}(\mathbf{k})$, $\varphi_{2\lambda}(\mathbf{k})$, $\varphi_{2\lambda}(\mathbf{k},\mathbf{k}^{\prime})$ can be obtained from the analytical structure of the Green function $G_{2}(\omega)=\int dtG_{2}(t)e^{i{\omega}t}$ in the frequency domain~\cite{shi16a}, where $G_{2}(t)=-i\langle{0}|\alpha_{2}(t)\alpha_{1}(t)\alpha_{1}^{\dagger}\alpha_{2}^{\dagger}|0\rangle\theta(t)$ in the time domain, and $\alpha_{1,2}\in(b_{1,2},a_{\mathbf{k}},a_{\mathbf{k}}^{\prime})$ (see~\ref{app:2ES1}).

The ground state energy corresponds to the smallest isolated pole of $G_{2}(\omega)$ in the first Riemann surface. As we shall discuss in Sec.~\ref{sec:2EGS}A, the ground state describes two repulsive polaritons localized around two different QEs, where the localization behavior is analyzed by the ground state wavefunctions in Sec.~\ref{sec:2EGS}A. We also construct a variational wavefunction and an effective Hamiltonian to describe the low energy physics in the two-excitation subspace.

Apart form the pole corresponding to the ground state, one can also find two additional isolated poles corresponding to higher excited states in the certain parameter regime (see Sec.~\ref{sec:2EDL:1} and~\ref{app:2ES1}). These two states can be illustrated in the following way: for a single QE with two photons, it has been demonstrated that the two photons can be both localized around the QE and form a two-excitation bound state~\cite{shi09a,shi16a,calajo16a}, which is referred to as the doublon state, schematically depicted in Fig.~\ref{fig1}(b). Here, the two higher excited states represent the symmetric and anti-symmetric superpositions of doublon states around different QEs. The properties of the doublon state is studied in Sec.~\ref{sec:2EDL:1}, where an effective hopping model for the doublon is derived.

\subsection{QE array}

Here, two new types of states appear compared to the single-excitation regime, i.e., the scattering state of two polaritons and the propagating doublon state.  For a system with two excitations in the QE array, the eigenstate at quasi-momentum $\mathbf{q}$ has the general form
\begin{eqnarray}
|\Psi_{2\lambda}(\mathbf{q})\rangle &=& \sum_{\mathbf{p}}f_{b}(\mathbf{p}) b_{\frac{\mathbf{q}}{2}+\mathbf{p}}^{\dagger} b_{\frac{\mathbf{q}}{2}-\mathbf{p}}^{\dagger} |0\rangle + \sum_{\mathbf{p,K}}f_{ba}(\mathbf{p,K}) b_{\frac{\mathbf{q}}{2}+\mathbf{p}}^{\dagger} a_{\frac{\mathbf{q}}{2}-\mathbf{p+K}}|0\rangle \nonumber \\
&+& \sum_{\mathbf{p,K,K^{\prime}}}f_{a}(\mathbf{p,K,K^{\prime}}) a_{\frac{\mathbf{q}}{2}+\mathbf{p+K}}^{\dagger}a_{\frac{\mathbf{q}}{2}-\mathbf{p+K}^{\prime}}^{\dagger}|0\rangle,\label{eq:2doub}
\end{eqnarray}
where the momenta $\mathbf{p}$ and $\mathbf{q}$ are restricted to the first Brillouin zone of the QE's sublattice. One can introduce the four-point Green function $G_{2}(\mathbf{q},t)=-i\langle{0}|\alpha_{2}(t)\alpha_{1}(t)\alpha_{1}^{\dagger}\alpha_{2}^{\dagger}|{0}\rangle\theta(t)$ in the time domain, where the operators $\alpha_{1}\in(b_{\mathbf{q}/2+\mathbf{p}},b_{\mathbf{q}/2+\mathbf{p}},a_{\mathbf{q}/2+\mathbf{p+K}})$ and $\alpha_{2}\in(b_{\mathbf{q}/2-\mathbf{p}},a_{\mathbf{q}/2-\mathbf{p+K}},a_{\mathbf{q}/2-\mathbf{p+K}^{\prime}})$. Its Fourier transform $G_{2}(\mathbf{q},\omega)=\int dtG_{2}(\mathbf{q},t)e^{i{\omega}t}$ gives the dispersion relation $E_{2\lambda}(\mathbf{q})$ of the state $|\Psi_{2\lambda }(\mathbf{q})\rangle $ in the band $\lambda$, the wavefunctions $f_{b}(\mathbf{p})$, $f_{ba}(\mathbf{p,K})$, and $f_{a}(\mathbf{p},\mathbf{K,K}^{\prime})$. The energy band structure in the two-excitation subspace can be determined from the analytical structure of $G_{2}(\mathbf{q},\omega)$ (see~\ref{app:2ES2}).

In Fig.~\ref{bandstructure}, we show the three lowest bands of the two excitation subspace for different sets of system parameters with 1D tight-binding dispersion relation.  The bands corresponding to the scattering of two polaritons are identified by the continuum in Fig.~\ref{bandstructure} (in gray shading), where the lowest band belongs to them. The scattering properties will be studied in detail in Sec.~\ref{sec:2EGS}, where an effective two-body Hamiltonian is established to describe the low energy dynamics of two polaritons. Between the scattering continuum, the single isolated band appears in the midgap, which corresponds to the propagating doublon, which can be described with an effective hopping model as we show in Section~\ref{sec:2EDL}.


\section{Two Excitations: Standard Polaritons \label{sec:2EGS}}


\subsection{Two-QE ground state \label{sec:2EGS:1}}

In this subsection, we study the ground state of the two excitation subspace when two QEs coupled to the photonic bath. In particular, i) we compute the exact ground state wavefunction; ii) we construct a variational ans\"{a}tz to reveal the properties of the ground state; and iii) we derive an effective Hamiltonian to describe the low-energy dynamics in the so-called arc region.

\begin{figure}[tbp]
\centering
\includegraphics[width=0.7\linewidth]{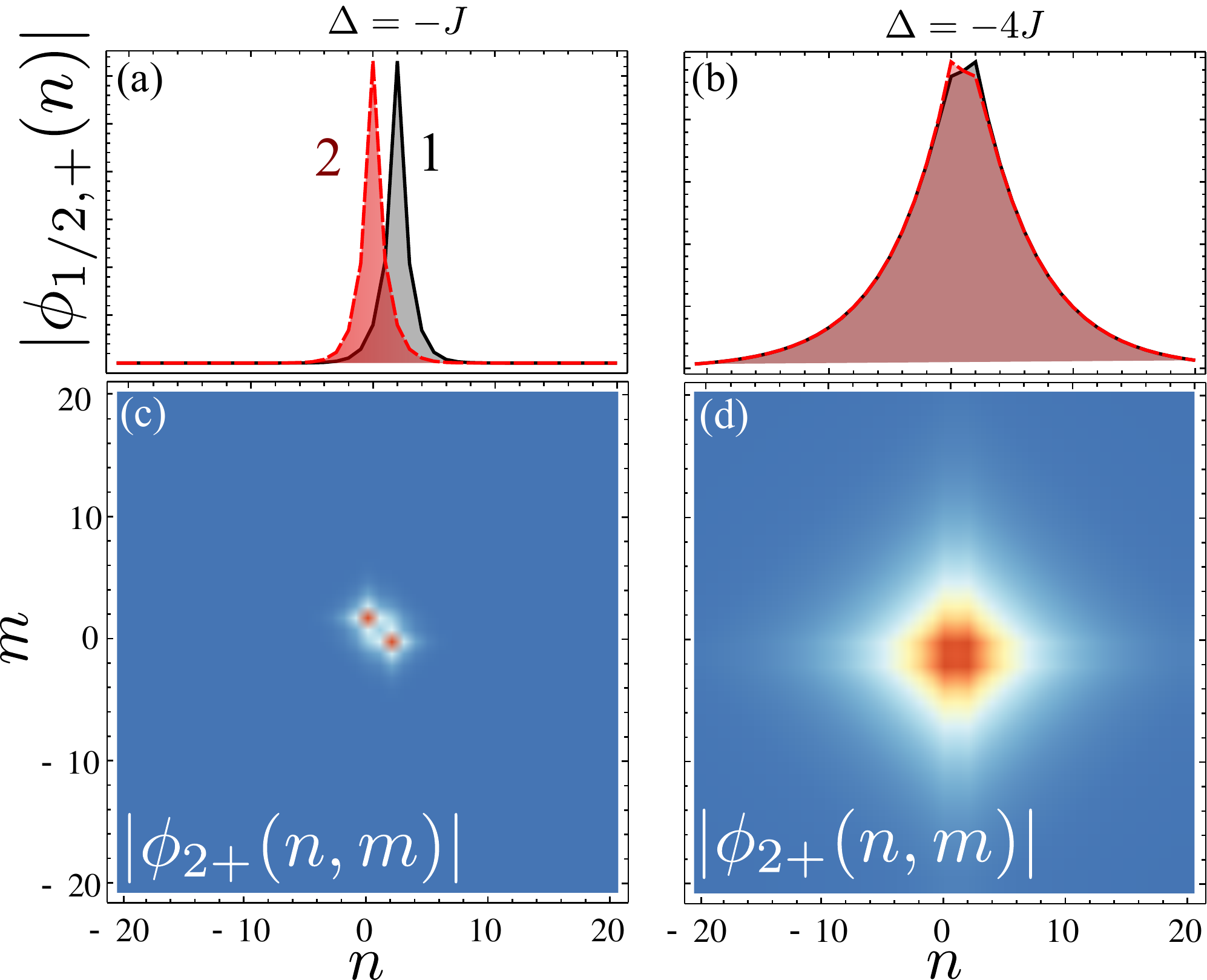}
\caption{The single-photon,  $\varphi_{j\rm{G}}(n)$, and two-photon wavefunctions, $\varphi_{2\rm{G}}(\mathbf{n},\mathbf{m})$, as defined in the text, for a $\cos(k)$ bath dispersion and a QE with detunings $\Delta/J=-1$ (a) and $\Delta/J=4$ (b), where $\Omega/J=1$ and $d=2$.}
\label{twoWf}
\end{figure}

The ground energy $E_{\rm{G}}$ is determined by the position of the isolated pole in the first Riemann surface of $G_{2}(\omega )$, more precisely, $E_{\rm{G}}$ is the smallest solution of the equation
\begin{equation}
\sum_{\sigma=\pm,\lambda\lambda^{\prime}}\frac{Z_{1\lambda}^{\sigma}Z_{1\lambda^{\prime}}^{\sigma}}{E_{\rm{G}}-E_{1\lambda }-E_{1\lambda^{\prime}}}=0,
\end{equation}
where $E_{1\lambda}$ and $Z_{1\lambda}^{\sigma}$ are defined in Sec.~\ref{TQE}. The ground state configuration is visualized by the wavefunctions $\varphi_{j\rm{G}}(\mathbf{k})$ and $\varphi_{2\rm{G}}(\mathbf{k},\mathbf{k}^{\prime})$ (their analytical expressions are given in~\ref{app:2ES1}). In the left and right panels of Fig.~\ref{twoWf}, we show the wavefunctions $\varphi_{j\rm{G}}(\mathbf{n})=\sum_{\mathbf{k}}\varphi_{j\rm{G}}(\mathbf{k})e^{i\mathbf{k\cdot n}}/\sqrt{N}$ and $\varphi_{2\rm{G}}(\mathbf{n},\mathbf{m})=\sum_{\mathbf{kk}^{\prime}}\varphi_{2\rm{G}}(\mathbf{k},\mathbf{k}^{\prime})e^{i\mathbf{k\cdot n}+i\mathbf{k}^{\prime}\mathbf{\cdot m}}/N$ in the coordinate space for $\Delta/J=-1$ and $\Delta/J=4$, respectively, where the distance $d=2$ and $\Omega/J=1$. For $\Delta/J=-1$, the ground state is mostly composed of QE excitations with hard-core interaction, thus the two excitations repulse each other and prefer to localize around two different QEs, as shown in Figs.~\ref{twoWf}a and~\ref{twoWf}c. For $\Delta/J=4$, the ground state is dominated by the free photons, and the single excitations localized around different QEs hybridize with each other, as shown in Figs.~\ref{twoWf}b and \ref{twoWf}d.

\begin{figure}[tbp]
\centering
\includegraphics[width=0.7\linewidth]{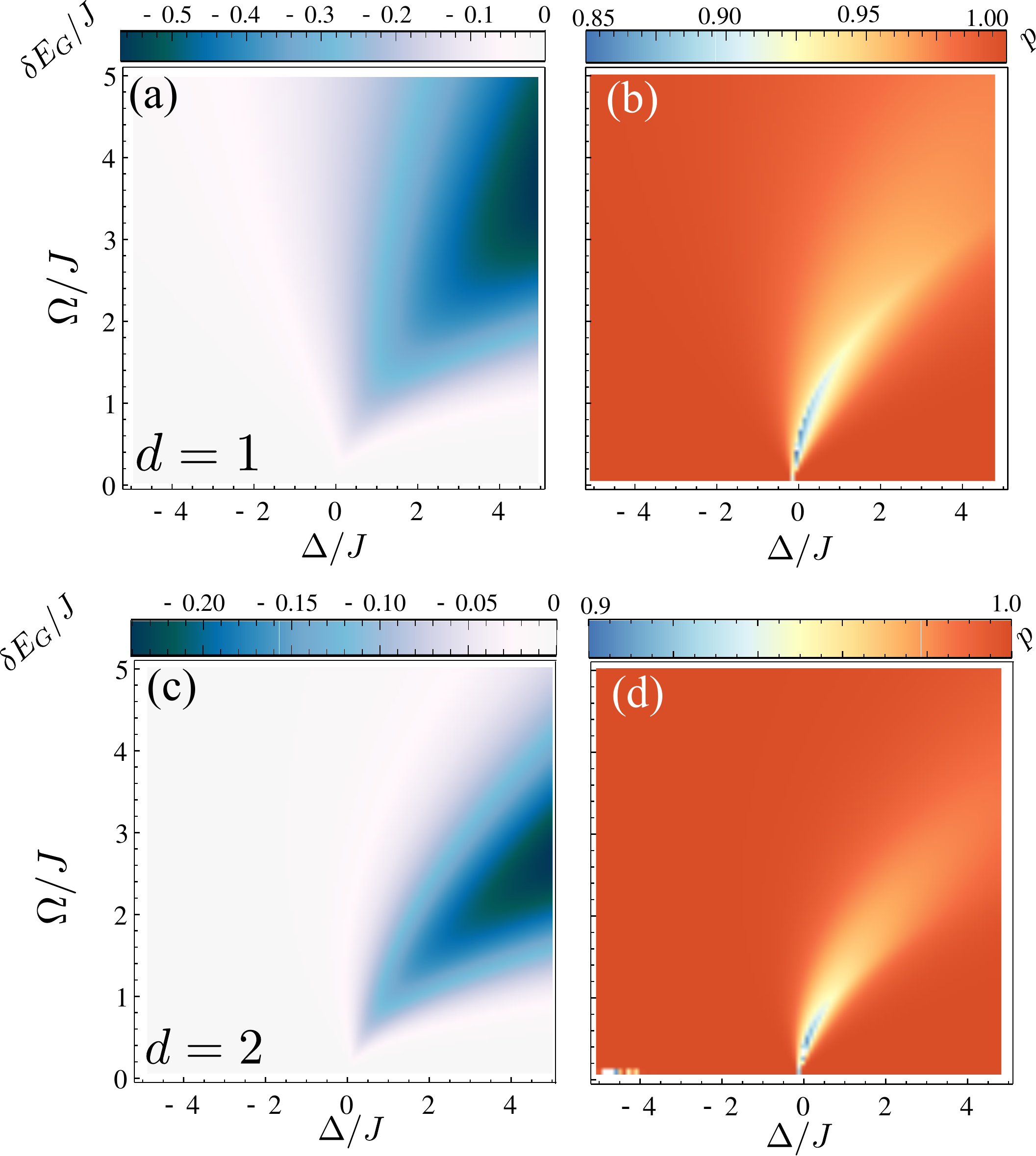}
\caption{The energy difference $\delta{E}_{\rm{G}}$ and the weight $p$, as defined in the main text, for $d=1$ (a)-(b) and $d=2$ (c)-(d).}
\label{twoGS}
\end{figure}

This repulsive interaction between the two excitations can also be identified by the energy difference $\delta E_{\rm{G}}\equiv E_{\rm{G}}-2E_{1B}$ the ground state energy $E_{\rm{G}}$ and the energy of the excitations localized around two individual QEs far apart from each other. In Figs.~\ref{twoGS}a and \ref{twoGS}c, we show $\delta E_{\rm{G}}$ in the $\Delta-\Omega$ plane for $d=1$ and $2$, respectively. In the arc region, the energy difference $\delta E_{\rm{G}}\sim 0$. This is because the two single-excitation bound states are strongly localized around different QEs, which suppresses their interaction. In the regime $\Delta >0$ and $\Omega/J{\ll}1$, the non-interacting photon excitations dominate in the bound state, and $\delta E_{\rm{G}}{\sim}0$. In the vicinity of the boundary $\Omega=\sqrt{2\Delta /d}$, where the anti-symmetric bound state $\beta_{-}^{\dagger}|0\rangle$ vanishes, there is a still a considerable probability for the QEs to be excited, while having a large overlap between the two single-excitation bound states. This induces the interaction between the two Wannier modes such that $\delta E_{\rm{G}}<0$, as shown by the dark (blue) regions in Figs.~\ref{twoGS}a and~\ref{twoGS}c.

The low energy dynamics of the single excitation subspace is governed by the effective Hamiltonian $H_{\rm{eff}}^{(1)}$ projected into the subspace spanned by the polariton modes $\beta_{\pm}^{\dagger}|0\rangle$. Therefore, we expect that the two-excitation ground state describing the two interacting polaritons is mostly composed of the low energy excitations $\beta_{\pm}^{\dagger{2}}|0\rangle/\sqrt{2}$. In Figs.~\ref{twoGS}c and~\ref{twoGS}d, we show the probability $p=\sum_{\sigma=\pm}\left| \langle{0}|\beta_{\sigma}^{2}|\Psi_{2\rm{G}}\rangle \right|^{2}/2$ to find two (anti-) symmetric excitations $\beta_{\pm}^{\dagger{2}}|0\rangle/\sqrt{2}$ in the exact ground state $|\Psi_{2\rm{G}}\rangle$. Here, we note that in the regime $\Delta>\Omega^{2}d/2$ the anti-symmetric mode $\beta_{-}^{\dagger}|0\rangle$ vanishes, and the probability is defined as $p=\left| \langle{0}|\beta_{+}^{2}|\Psi_{2\rm{G}}\rangle \right|^{2}/2$, namely, only the symmetric mode is taken into account. The large probability $p>0.85$ even in the non-Markovian regime shows that the excitations $\beta_{\pm}^{\dagger{2}}|0\rangle/\sqrt{2}$ dominate the ground state, which indicates that the low energy dynamics in the single and two-excitation subspaces can be described by some interacting Hamiltonian for the polariton modes $\beta_{\pm}^{\dagger}|0\rangle$.

A remarkable feature of the two excitation ground state is that the first order correlation matrix $M_{ij}=\langle \alpha_{i}^{\dagger}\alpha_{j} \rangle$, for operators $\alpha_{j}=b_{1,2},a_{\mathbf{k}}$, and the two-photon wavefunction $\varphi_{2\rm{G}}(\mathbf{k},\mathbf{k}^{\prime})$ at most have two dominating singular values in the whole $\Delta-\Omega$ plane. This fact inspires us to construct a variational ans\"{a}tz $|\Psi_{2\rm{v}}\rangle=\mathcal{N}_{2}^{-1/2}\left(\gamma_{+}^{\dagger{2}}-\gamma_{-}^{\dagger{2}}\right)|0\rangle/2$ for the ground state by two deformed single-excitation modes defined by:
\begin{equation}
\gamma_{\pm }^{\dagger}=\frac{1}{\sqrt{2}} \left( b_{1}^{\dagger}{\pm} b_{2}^{\dagger}\right) + v_{\pm}\sum_{\mathbf{k}}\varphi_{\pm}(\mathbf{k})a_{\mathbf{k}}^{\dagger},
\end{equation}
where $v_{\pm}$ are real numbers, $\varphi_{\pm}(\mathbf{k})$ are orthonormal wavefunctions, and the normalization factor $\mathcal{N}_{2}=\sum_{\sigma}(1+v_{\sigma}^{2})^{2}/2$.

Under the normalization constraint $\sum_{\mathbf{k}}|\varphi_{\sigma}(\mathbf{k})|^{2}=1$, the minimization of the variational energy $E=\langle\Psi_{2\rm{v}}|H|\Psi_{2\rm{v}}\rangle$ with respect to $\varphi_{\sigma}(\mathbf{k})$ indicates that the variational function should have the form
\begin{equation}
\varphi_{\sigma}(\mathbf{k})=\frac{1}{\sqrt{\mathcal{N}_{\sigma }N}}\frac{\eta_{\mathbf{k}\sigma}}{e_{\sigma }-\varepsilon_{\mathbf{k}}},
\end{equation}
where $e_{\sigma}$ is a variational parameter, $\eta_{\mathbf{k}\sigma}=(1+{\sigma}e^{-i\mathbf{k{\cdot}d}})/\sqrt{2}$, and the normalization factor
\begin{equation}
\mathcal{N}_{\sigma}=\frac{1}{N}\sum_{\mathbf{k}}\frac{1+\sigma\cos\mathbf{k{\cdot}d}}{(e_{\sigma }-\varepsilon_{\mathbf{k}})^{2}}.
\end{equation}
In terms of the variational parameters $v_{\sigma}$ and $e_{\sigma}$, the variational ground state energy can be written as
\begin{eqnarray}
E &=& \frac{1}{\mathcal{N}_{2}} \Bigg\{ 2\Delta + \sum_{\sigma} \Big[\Delta v_{\sigma}^{2}+v_{\sigma}^{2}(1+v_{\sigma}^{2})e_{\sigma} \nonumber \\
&+& \frac{1}{\sqrt{\mathcal{N}_{\sigma}}} v_{\sigma}(1+v_{\sigma}^{2}) \left( 2\Omega-\frac{v_{\sigma }}{\sqrt{\mathcal{N}_{\sigma}}}\right) I_{\sigma}(e_{\sigma}) \Big] \Bigg\},
\end{eqnarray}
where
\begin{equation}
I_{\sigma }(e_{\sigma})=\frac{1}{N}\sum_{\mathbf{k}}\frac{1+\sigma \cos \mathbf{k{\cdot}d}}{e_{\sigma}-\varepsilon_{\mathbf{k}}}.
\end{equation}

\begin{figure}[tbp]
\centering
\includegraphics[width=0.65\linewidth]{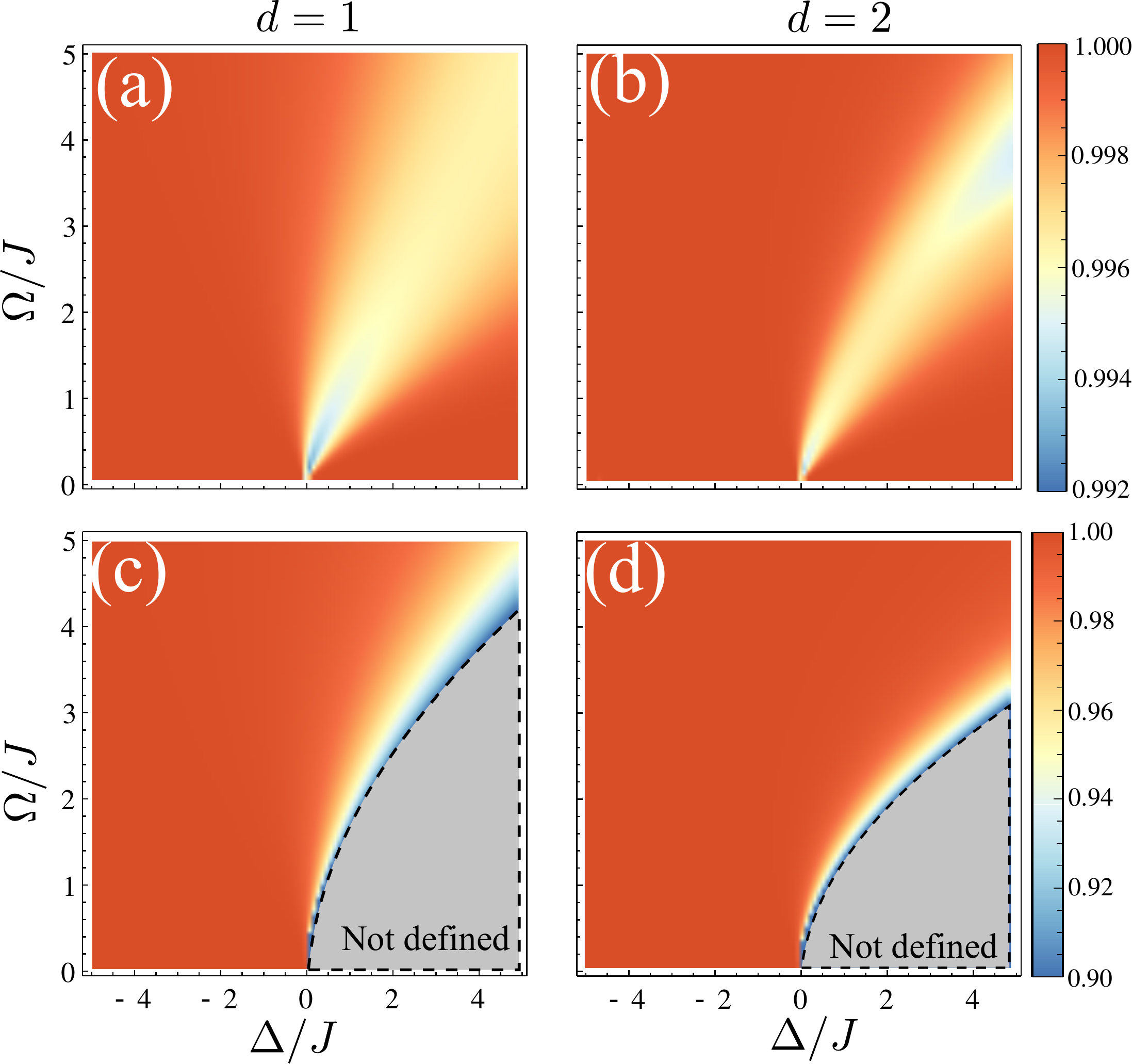}
\caption{The overlap between the variational state and the exact ground state for (a) $d=1$ and (b) $d=2$. The overlap between the deformed single-particle mode $\gamma_{\rm{N},-}^{\dagger}|0\rangle$ and the bared anti-symmetric mode $\beta_{-}^{\dagger}|0\rangle$ for (c) $d=1$ and (d) $d=2$.}
\label{pv}
\end{figure}

One can minimize the ground state energy with respect to $v_{\sigma}$ and $e_{\sigma}$ to find the optimal variational state $|\Psi_{\rm{v}2}\rangle$. Figures~\ref{pv}a and~\ref{pv}b show the overlap $p_{\rm{v}}$ between $|\Psi_{\rm{v}2}\rangle$ and the exact ground state for 1D tight-binding dispersion relation at $d=1,2$. The fact that $p_{\rm{v}}>0.98$ in all parameter regimes underscores the accuracy of our variational state $|\Psi_{\rm{v}2}\rangle$.

If we define the normalized symmetric and anti-symmetric modes $\gamma_{\rm{N},\pm}=\gamma_{\pm}/\sqrt{1+v_{\pm}^{2}}$ of the variational state, we can calculate the overlap between these modes and the single-particle ones,  $\beta_{\pm}^{\dagger}|0\rangle$, given by $p_{\pm}=\left| \langle{0}| \beta_{\pm}\gamma_{\rm{N},\pm}^{\dagger} |0\rangle \right|$. The symmetric mode $\gamma_{\rm{N},+}^{\dagger}|0\rangle$ only deviates slightly from the bare single-particle mode $\beta_{+}^{\dagger}|0\rangle$, i.e., $p_{+}>0.99$, in the whole parameter space, including the non-Markovian regimes (not shown). On the contrary, the mode $\gamma_{\rm{N},-}^{\dagger}|0\rangle $ is very different from the anti-symmetric single-particle mode $\beta_{-}^{\dagger}|0\rangle$ in the vicinity of the boundary $\Delta=\Omega^{2}d/2$, as shown in Figs.~\ref{pv}c and \ref{pv}d for $d=1,2$. Furthermore, in the regime $\Delta>\Omega^{2}d/2$, the anti-symmetric bound state vanishes and the $\gamma_{\rm{N},+}^{\dagger{2}}|0\rangle/\sqrt{2}$ dominates the ground state. This is why the probability $p$, which measures the overlap between the exact and the variational wavefunction, is still very large in this non-Markovian regime.

In the so-called arc region, $p_{\pm}{\sim}1$ and $v_{+}=v_{-}$ indicates that the ground state $|\Psi_{2\rm{v}}\rangle\sim\beta_{1}^{\dagger}\beta_{2}^{\dagger}|0\rangle$, where the small component of the double occupation states $\beta_{j=1,2}^{\dagger{2}}|0\rangle$ in $|\Psi_{2\rm{v}}\rangle$ shows the hard-core nature of $\beta_{j}^{\dagger}|0\rangle$. As a result, we can construct the effective spin model
\begin{eqnarray}
H_{\rm{eff}} &=& t_{\rm{eff}}(\tilde{\sigma}_{1}^{+}\tilde{\sigma}_{2}^{-} + \tilde{\sigma}_{2}^{+}\tilde{\sigma}_{1}^{-}) + \frac{1}{4} J_{z}\tilde{\sigma}_{1}^{z}\tilde{\sigma}_{2}^{z} + \frac{1}{4}E_{\rm{G}}(\tilde{\sigma}_{1}^{z}+\tilde{\sigma}_{2}^{z})+\frac{1}{4}J_{z}+E_{0}
\end{eqnarray}
for the single and two excitations subspaces, where $J_{z}=E_{\rm{G}}-2E_{0}$ and the spin operators $\tilde{\sigma}_{j=1,2}^{-}$ denote the local Wannier modes $\tilde{\beta}_{1,2}$. In the Markovian regime the spin operators $\tilde{\sigma}_{j}^{-}$ become the bare QE transition operator $\sigma_{1,2}^{-}$, while in the strong coupling regime $\tilde{\sigma}_{j}^{-}$ denotes the annihilation operator of the strongly hybridized polariton mode.

\subsection{Many QEs \label{sec:2EGS:2}}

For two QEs, the two-excitation ground state can be described by two interacting single-excitations $\beta_{\pm}^{\dagger{2}}|0\rangle/\sqrt{2}$. In this subsection, we investigate the scattering of two excitations in the periodic QE array coupled to the photonic bath. As discussed in Sec.~\ref{sec:GF}, the two-excitation spectrum is obtained by the four-point connected Green function (see~\ref{app:2ES2}), which displays a band structure due to the PBC imposed. To study the low energy scattering of two individual polaritons, we derive the energy-dependent two-body interaction in the whole parameter plane using the Feshbach treatment~\cite{pethickbook02a}. We demonstrate that the effective interaction strength between two polaritons can be tuned by the detuning $\Delta $ and the Rabi frequency $\Omega$. In particular, the hardcore nature of two excitations justifies the validity of the spin model in the arc region.

\begin{figure}[tbp]
\centering
\includegraphics[width=0.7\linewidth]{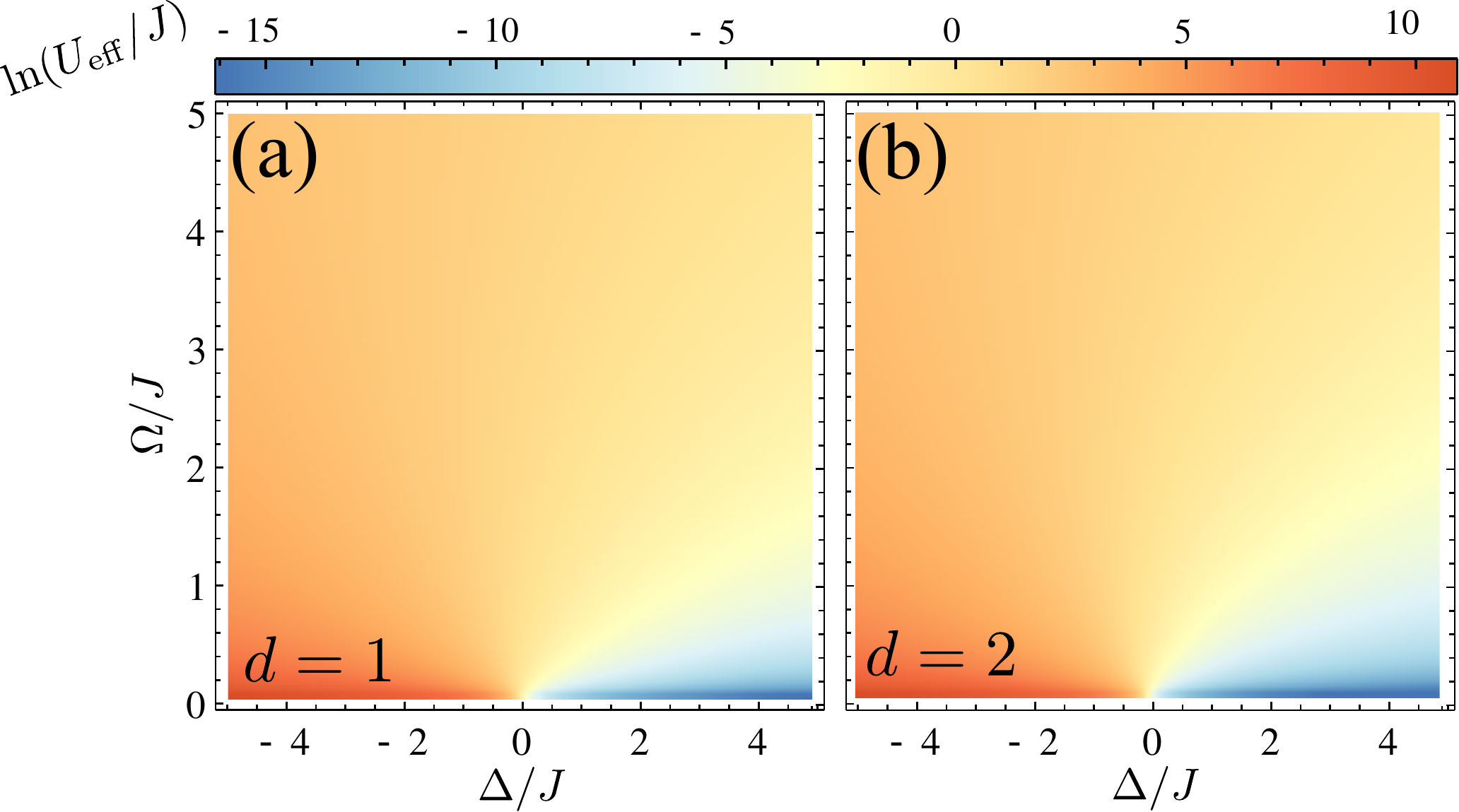}
\caption{The low-energy two-body interaction in the parameter plane, where (a) $d=1$ and (b) $d=2$.}
\label{Ueff}
\end{figure}

Let us first derive the effective Hamiltonian describing the lowest band of two-excitation spectrum using the Feshbach treatment. For the incident state $\beta_{\mathbf{p}}^{\dagger}\beta_{\mathbf{q}-\mathbf{p}}^{\dagger}|0\rangle$ of two polaritons with total momentum $\mathbf{q}$ and energy $E$, the scattering state $|\Psi_{2\rm{sc}}(\mathbf{q})\rangle=|\Psi_{2P}(\mathbf{q})\rangle+|\Psi_{2Q}(\mathbf{q})\rangle $ can be written as the superposition of $|\Psi_{2P}(\mathbf{q})\rangle=P|\Psi_{2\rm{sc}}(\mathbf{q})\rangle $ and $|\Psi_{2Q}(\mathbf{q})\rangle=Q|\Psi_{2\rm{sc}}(\mathbf{q})\rangle$, where $P$ and $Q=1-P$ are the projectors into the lowest and higher bands, i.e., the ``open" and ``closed" channels in the Feshbach resonance.

By eliminating the higher energy band in the Schr\"{o}dinger equation, the component $\left\vert \Psi _{2P}(\mathbf{q})\right\rangle $ in the lowest channel obeys the secular equation
\begin{equation}
H_{\rm{eff}}(E)|\Psi_{2P}(\mathbf{q})\rangle = E |\Psi_{2P}(\mathbf{q})\rangle.
\end{equation}
The effective Hamiltonian $H_{\rm{eff}}(E)=H_{\rm{eff}}^{(1)}+PH_{\rm{int}}(E)P$ describes the scattering of two excitations in the lowest band by
\begin{equation}
H_{\rm{int}}(E) = \frac{1}{2N_{b}}\sum_{\mathbf{p}_{1}\mathbf{p}_{2}\mathbf{q}} U_{\rm{int}}(\mathbf{q,}E) b_{\mathbf{p}_{1}}^{\dagger} b_{\mathbf{q}-\mathbf{p}_{1}}^{\dagger} b_{\mathbf{q}-\mathbf{p}_{2}} b_{\mathbf{p}_{2}},
\end{equation}
where the effective two-body interaction
\begin{equation}
U_{\rm{int}}(\mathbf{q},E) = -\left[ \frac{1}{N_{b}}\sum\nolimits_{\mathbf{p,}\lambda\lambda^{\prime}}^{\prime} \frac{ Z_{1\lambda}(\mathbf{p})Z_{1\lambda^{\prime}}(\mathbf{q}-\mathbf{p}) }{ E-E_{1\lambda}\mathbf{p})-E_{1\lambda^{\prime}}(\mathbf{q}-\mathbf{p}) } \right]^{-1}
\end{equation}
depends on the incident energy $E$, and the summation does not include the contribution from the lowest band. The component 
\begin{equation}
|\Psi_{2Q}(\mathbf{q})\rangle = \frac{1}{E-H_{0}}QH_{\rm{int}}(E)|\Psi_{2P}(\mathbf{q})\rangle
\end{equation}
in the closed channel follows from the Schr\"{o}dinger equation, which is a bound state of two excitations in the higher energy bands. Thus, the closed channel component will not contribute to the scattering wavefunction in the asymptotic limit.

We characterize the scattering process of two low-energy polaritons at the band bottom by the effective interaction $U_{\rm{eff}}=Z_{1}^{2}(0)U_{\rm{eff}}(0,E_{0})$, which we plot in Fig.~\ref{Ueff} for $d=1,2$ in the $\Delta-\Omega$ plane. In the $\Delta>0$ regime, the polariton mode $\beta_{\mathbf{p}}^{\dagger }|0\rangle $ is mostly composed of bath photons for small $\Omega $, so the interaction strength $U_{\rm{eff}}$ is extremely softened. As $\Omega $ increases, the weight of QE excitation in the mode $\beta_{\mathbf{p}}^{\dagger}|0\rangle$ becomes larger, and the effective interaction $U_{\rm{eff}}$ increases accordingly. This means a wide range of $U_{\rm{eff}}$ can be achieved by tuning the Rabi frequency $\Omega$ and $\Delta$.

In the Markovian regime, the single-particle weight $Z_{1}{\sim}1$ and a large gap (compared with the bandwidth) separates the lowest band and the higher energy bands, which result in the divergent interaction $U_{\rm{eff}}$ and the small component $||\Psi _{2Q}(\mathbf{q})\rangle|^{2}$ in the closed channel. Thus, the low-energy dynamics can be described by the effective spin model $H_{\rm{eff}}=\sum_{jj^{\prime}}t_{j-j^{\prime}}\tilde{\sigma}_{j}^{+}\tilde{\sigma}_{j^{\prime }}^{-}$ projected in the lowest band, where the spin operator $\tilde{\sigma}_{j}^{-}\sim\sigma_{j}^{-}$ can be approximated by the bared QE transition operator. In the strong coupling regime, $Z_{1}\sim 1/2$ and the two-body interaction $U_{\rm{eff}}$ between strongly hybridized polaritons is divergent. Therefore, the spin model is also valid in this region, and $\tilde{\sigma}_{j}^{-}\sim\beta_{j}$ denotes the local Wannier mode with the strong on-site interaction. 

\section{Two Excitations: Doublons \label{sec:2EDL}}


\subsection{Two-QE doublon state \label{sec:2EDL:1}}

\begin{figure}[tbp]
\centering
\includegraphics[width=0.7\linewidth]{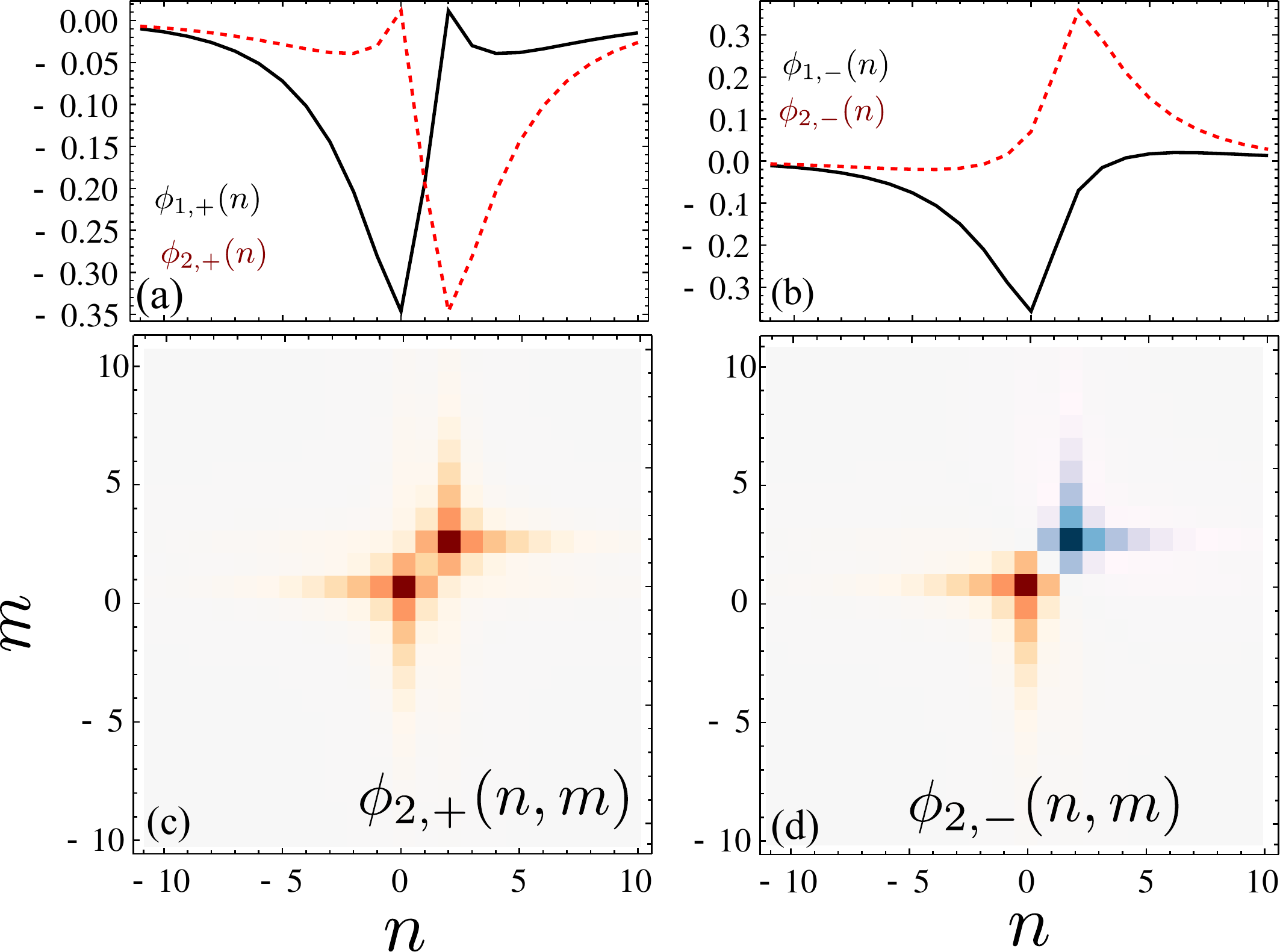}
\caption{The wavefunctions of the symmetric and anti-symmetric bound states are shown in the left and right panels, where $d=2$, $\Omega/J=2$, and $\Delta/J=-1$. The single-photon wavefunction $\protect\varphi_{j\pm}(\mathbf{n})$ and the two-photon wavefunction $\varphi_{2\pm}(\mathbf{n},\mathbf{m})$ are shown in the upper and lower rows.}
\label{twoEX}
\end{figure}

As it occurred for the single-photon bound states, only for a given region of the $\Delta-\Omega$ parameter space, two bound states $|\Psi_{2\pm}\rangle$ with higher energies $E_{2\pm}$ than $E_{\rm{G}}$ appear in the spectrum. The exact condition for the existence of two bound states is determined via the Green function approach in~\ref{app:2ES1}. The localization behavior of two bound states can be understood by inspecting their wavefunctions $\varphi_{j\pm}(\mathbf{n})$ and $\varphi_{2\pm}(\mathbf{n},\mathbf{m})$, which we plot in Fig..~\ref{twoEX}, for $d=2$, $\Omega/J=2$, and $\Delta/J=-1$. There, we observe that the two bound states are the symmetric and anti-symmetric superpositions of the two-photon bound states~\cite{shi16a} localized around different QEs.

When both doublons exist an effective hopping model for them can be constructed, which reads:
\begin{equation}
H_{\rm{doublon}} = \mu_{D}\sum_{j=1,2}d_{j}^{\dagger}d_{j}+t_{D}(d_{1}^{\dagger}d_{2}+d_{2}^{\dagger }d_{1}),
\label{Hdoublon}
\end{equation}
where $d^\dagger_l$ describes the generation of the doublon Wannier mode $(|\Psi_{2+}\rangle-(-1)^{l}|\Psi_{2-}\rangle)/\sqrt{2}$ localized around the $l$-th QE with the chemical potential $\mu_{D}=(E_{2+}+E_{2-})/2$ and the effective hopping strength $t_{D}=(E_{2+}-E_{2-})/2$.

\begin{figure}[tbp]
\centering
\includegraphics[width=0.7\linewidth]{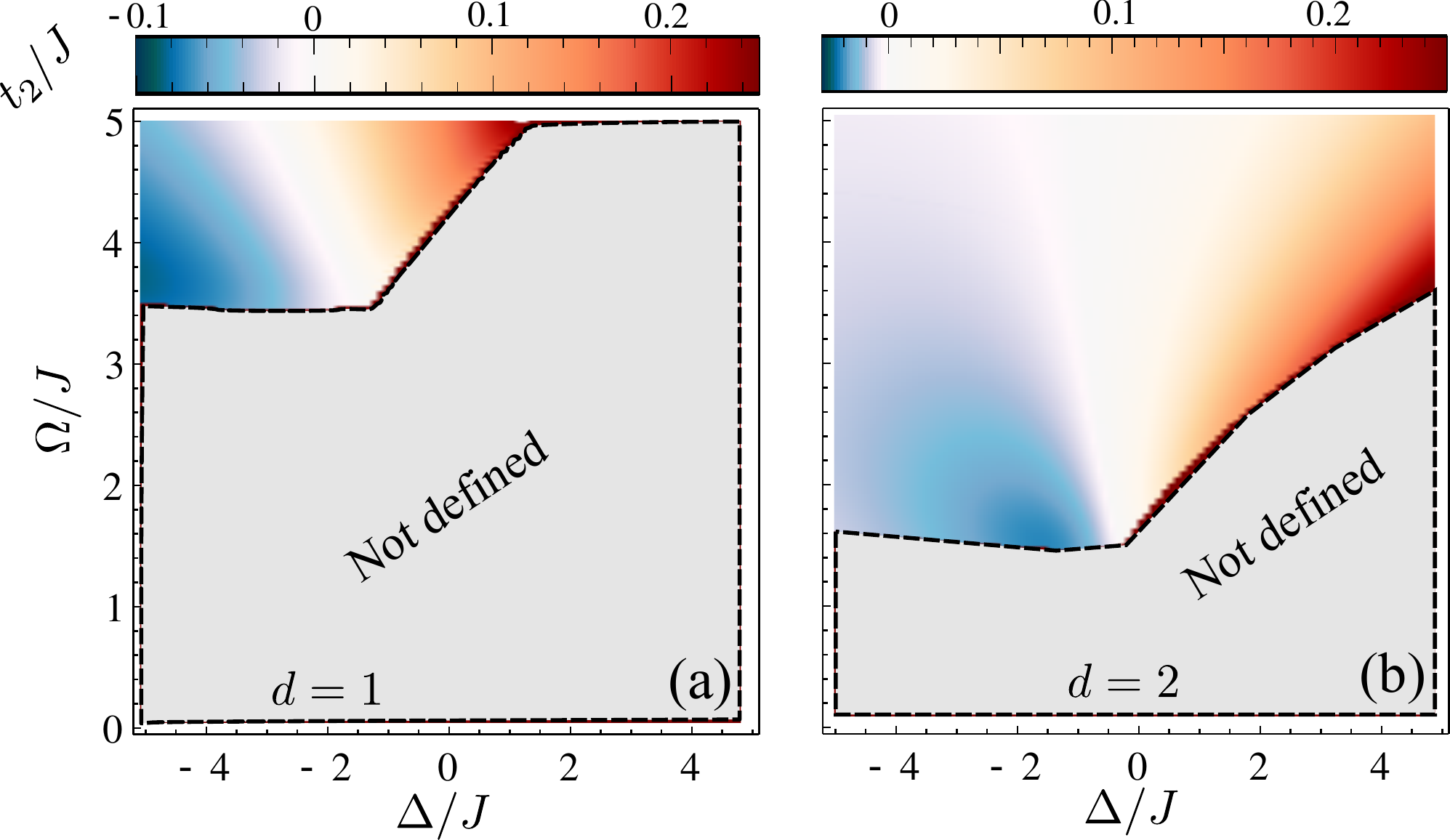}
\caption{The effective hopping $t_{D}$ for $d=1$ (a) and $d=2$ (b).}
\label{twot2}
\end{figure}

In Fig.~\ref{twot2} we plot the hopping strength $t_{D}$ in the $\Delta-\Omega$ plane for $d=1,2$. For fixed distance $d$ and small Rabi coupling, the large localization length of the doublon Wannier modes gives rise to a strong hybridization between them and a large energy level splitting $2t_{D}$. As a result, the two doublon states vanish by merging into the continuum of scattering band. As the Rabi frequency or the distance $d$ increases, the localization length is shorter than $d$, and the overlap between two doublon Wannier modes becomes smaller. Therefore, the energy level splitting $2t_{D}$ is reduced, so one can find two doublon states and define the effective hopping strength $t_{D}$. This intuitive picture also explains why the regime where these bound states vanish shrinks as the distance $d$ increases, as shown in Fig.~\ref{twot2}(b).

\subsection{Many QEs \label{sec:2EDL:2}}

For many QEs, we already saw in Fig.~\ref{bandstructure},  that a doublon band ($E_{2B}(\mathbf{q})$, in red triangles) appears in the midgap of the scattering bands, whose dispersion depends strongly on the parameter regime. For example, for $\Delta/J=1$ and $\Omega /J=1$ in Fig.~\ref{bandstructure}f, the doublon band has a visible curvature which implies that the doublons have a large hopping strength, whereas for $\Delta /J$ decreases to $-1$ in Fig.~\ref{bandstructure}d, the doublon band is very flat which tell us that the doublons are mostly localized.

\begin{figure}[tbp]
\centering
\includegraphics[width=0.7\linewidth]{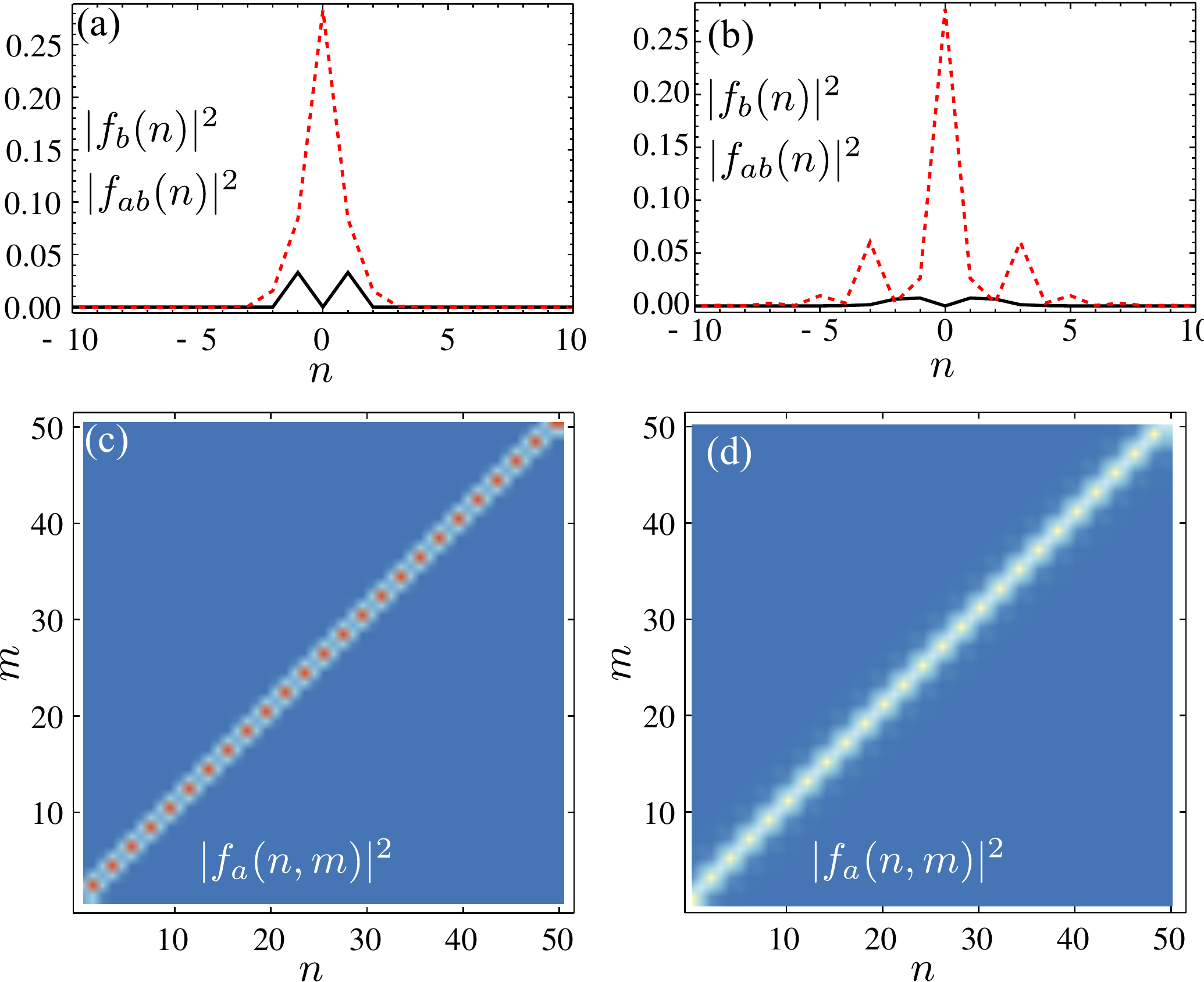}
\caption{The left and right panels display the wavefunctions of the doublon states with momentum $q=0$ in the first two doublon bands, respectively, where $d=2$, $\Delta/J=0$ and $\Omega/J=2$. In the first row, the wavefunctions $|f_{b}(\mathbf{n})|^{2}$ and $|f_{ab}(\mathbf{m,m})|^{2}$ are shown by the solid blue curves and dashed red curves, where the QE excitation in $f_{ab}(\mathbf{m,m})$ is set at the origin. The wavefunctions $\left\vert f_{a}(\mathbf{n},\mathbf{m})\right\vert ^{2}$ are shown in the second row.}
\label{W2}
\end{figure}

To gain more intuition of the features of this doublon band, we plot the coordinate space structure of the doublon in Fig.~\ref{W2} using the Fourier transforms $f_{b}(\mathbf{r})$, $f_{ba}(\mathbf{r,m})$, and $f_{a}(\mathbf{r},\mathbf{n,m})$ of the functions  $f_{b}(\mathbf{p})$, $f_{ba}(\mathbf{p,K})$, and $f_{a}(\mathbf{p},\mathbf{K,K}^{\prime})$ defined in Eq.~\ref{eq:2doub}. Here, $f_{b}(\mathbf{r})$ is the amplitude of having two excited QEs separated by a distance $\mathbf{r}$, $f_{ba}(\mathbf{r,m})$ is the amplitude of finding one photon at the position $\mathbf{m}$ and one excited QE separated from the photon by a distance $\mathbf{r}$, and $f_{a}(\mathbf{r},\mathbf{n,m})$ is the amplitude of detecting two photons at positions $\mathbf{n}$ and $\mathbf{m}$. In Fig.~\ref{W2}, we observe that for $d=2$, $\Delta/J=0$ and $\Omega/J=2$, the square norms $|f_{b}(\mathbf{r})|^{2}$, $|f_{ab}(\mathbf{r,m})|^{2}$, and $|f_{a}(\mathbf{r},\mathbf{n,m})|^{2}$ of the state with $\mathbf{q}=0$ in the first two isolated bands display that the two polaritons attract each other to form a propagating doublon. One can also see that the two polaritons are tightly bound in the lowest doublon band but loosely bound in the higher doublon band.

From the doublon state, $|\Psi_{2B}(\mathbf{q})\rangle$, and its dispersion relation $E_{2B}(\mathbf{q})$, we can construct the effective doublon Hamiltonian as follows:
\begin{equation}
H_{\rm{doublon}} = \sum_{\mathbf{n,m}} t_{\mathbf{n}-\mathbf{m}}^{D} d_{\mathbf{n}}^{\dagger} d_{\mathbf{m}},
\label{Hd}
\end{equation}
which describes the hopping of the doublon mode $d_{\mathbf{n}}^{\dagger}|0\rangle$. In the single doublon space, $d_{\mathbf{n}}=|0\rangle\langle\Psi_{2B}(\mathbf{n})|$ is defined by the Wannier state $|\Psi_{2B}(\mathbf{n})\rangle = \sum_{\mathbf{q}} e^{-i\mathbf{q}{\cdot}\mathbf{n}} |\Psi_{2B}(\mathbf{q})\rangle/\sqrt{N_{b}}$ localized around QE at the position $\mathbf{n}$. The effective hopping strength of the doublon is
\begin{equation}
t_{\mathbf{n}-\mathbf{m}}^{D}=\frac{1}{N_{b}}\sum_{\mathbf{q}}E_{2B}(\mathbf{q})e^{i\mathbf{q}{\cdot}(\mathbf{n-m})},
\end{equation}
where the NN hopping strength $t_{\mathbf{n-m}=1}^{D}$ is $t_{D}$ (the effective hopping strength in the two-QE case). We note that in the dilute gas limit, where the density of doublons ${\ll}1$, the effective hopping model $H_{\rm{doublon}}$ is still valid.

\section{Many Excitations \label{sec:manyexcitation}}

In this Section, we study the spectrum for the $N$-excitation subspaces (with $N>2$), focusing on the QE-array situation. In the first subsection, we apply the Green-function approach to characterize analytically the emergent dynamics in the $N=3$ subspace, finding a continuum band that describe two types of scattering processes, i.e., the scattering of three individual polaritons and that between one polariton and one doublon. There exists also an isolated band in which the three excitations form a bound state (referred to as the \textit{triplon} state) and co-propagate along the QE array.

Even though the Green-Function method can be in principle extended for larger excitation number, it becomes very challenging due to the emergence of many topologically inequivalent Feynman diagrams. Thus, in the second subsection we use the intuition developed by in the previous sections to numerically characterize the ground state properties of the system using DMRG~\cite{white92a,schollwock11a}. In particular, we will be able to show that one can go from a regime where the system behaves as a Mott-insulator to a superfluid behaviour, just by tuning the system parameters.

\subsection{Three excitations}

\begin{figure}[tbp]
\centering
\includegraphics[width=0.7\linewidth]{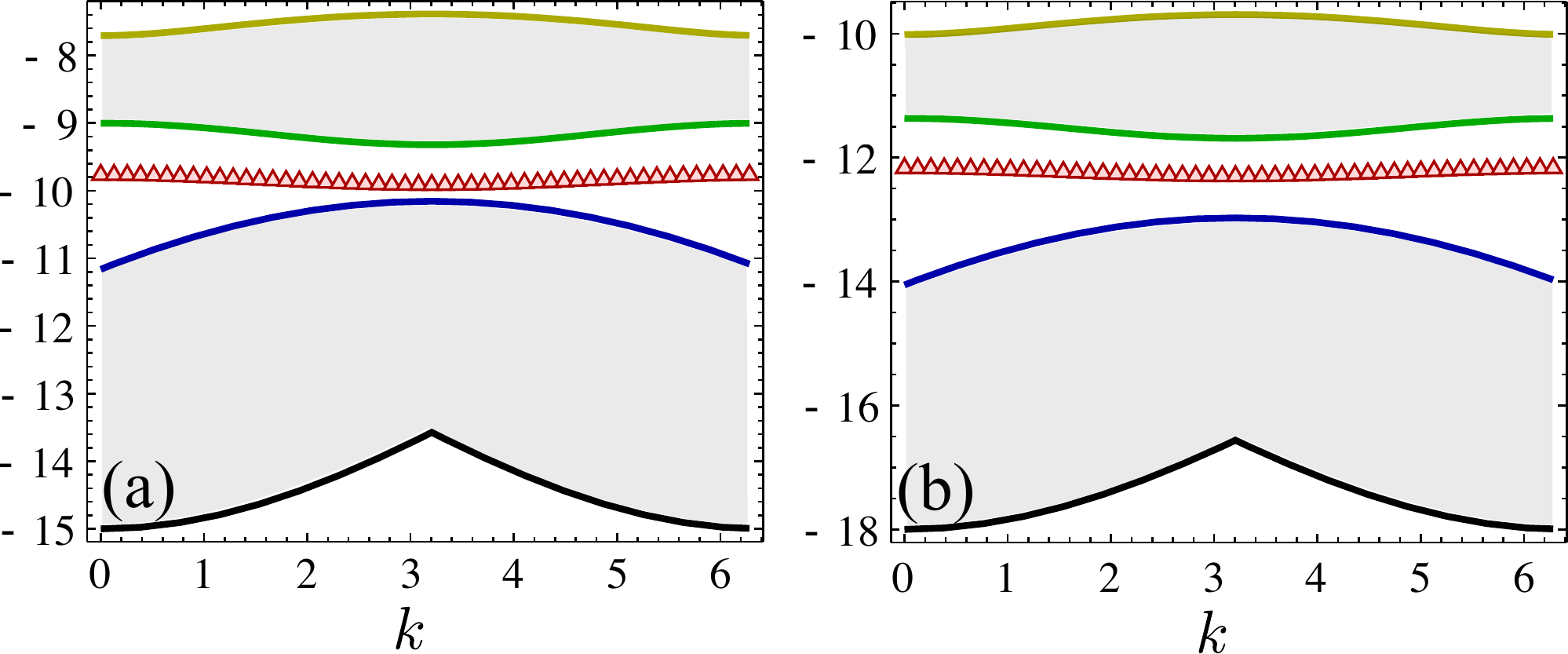}
\centering
\caption{The first three bands in the three-excitation subspace, where $\Delta=0$ and $d=1$. (a) $\Omega/J=5$; (b) $\Omega/J=6$.}
\label{E3B}
\end{figure}

In this subsection, we study the properties of three excitations using the Green function method (see~\ref{app:3ES}). In Fig.~\ref{E3B}, we show the band structure for three excitations for a situation with $\Delta=0$ and lattice spacing $d=1$. The lowest continuum band describes the scattering between three individual polaritons, while the second continuum band is composed of the scattering states between one polariton and one doublon. In the three-polariton scattering band, the two-body interaction between polaritons can be tuned by $\Omega$ and $\Delta$, as shown in Sec.~\ref{sec:2EGS}. With relatively large Rabi frequencies $\Omega/J=5,6$, the midgap opens between the two lowest scattering bands, and a triplon band (denoted by the red triangles) with dispersion relation $E_{3B}(\mathbf{q})$ appears. In the triplon band, the explicit analytic form of the triplon state $|\Psi_{3B}(\mathbf{q})\rangle$ reads:
\begin{eqnarray}
  \fl |\Psi_{3B}(\mathbf{q})\rangle = \sum_{\mathbf{k}_{1}\mathbf{k}_{2}} f_{b}(\mathbf{k}_{1},\mathbf{k}_{2}) b_{\mathbf{k}_{1}}^{\dagger} b_{\mathbf{k}_{2}}^{\dagger} b_{\mathbf{q}-\mathbf{k}_{1}-\mathbf{k}_{2}}^{\dagger} |0\rangle + \sum_{\mathbf{k}_{1}\mathbf{k}_{2}\mathbf{K}} f_{bba}(\mathbf{k}_{1},\mathbf{k}_{2},\mathbf{K}) b_{\mathbf{k}_{1}}^{\dagger} b_{\mathbf{k}_{2}}^{\dagger} a_{\mathbf{q}-\mathbf{k}_{1}-\mathbf{k}_{2},\mathbf{K}}^{\dagger} |0\rangle \nonumber \\
\fl + \sum_{\mathbf{k}_{1}\mathbf{k}_{2}\mathbf{K}_{2}\mathbf{K}_{3}} f_{baa}(\mathbf{k}_{1},\mathbf{k}_{2},\mathbf{K}_{2},\mathbf{K}_{3}) b_{\mathbf{k}_{1}}^{\dagger} a_{\mathbf{k}_{2},\mathbf{K}_{2}}^{\dagger} a_{\mathbf{q}-\mathbf{k}_{1}-\mathbf{k}_{2},\mathbf{K}_{3}}^{\dagger} |0\rangle \nonumber \\
\fl+ \sum_{\mathbf{k}_{1}\mathbf{k}_{2}\mathbf{K}_{1}\mathbf{K}_{2}\mathbf{K}_{3}} f_{a}(\mathbf{k}_{1},\mathbf{k}_{2},\mathbf{K}_{1},\mathbf{K}_{2},\mathbf{K}_{3}) a_{\mathbf{k}_{1},\mathbf{K}_{1}}^{\dagger}a_{\mathbf{k}_{2},\mathbf{K}_{2}}^{\dagger}a_{\mathbf{q}-\mathbf{k}_{1}-\mathbf{k}_{2},\mathbf{K}_{3}}^{\dagger} |0\rangle.
\end{eqnarray}

\begin{figure}[tbp]
\centering
\includegraphics[width=0.7\linewidth]{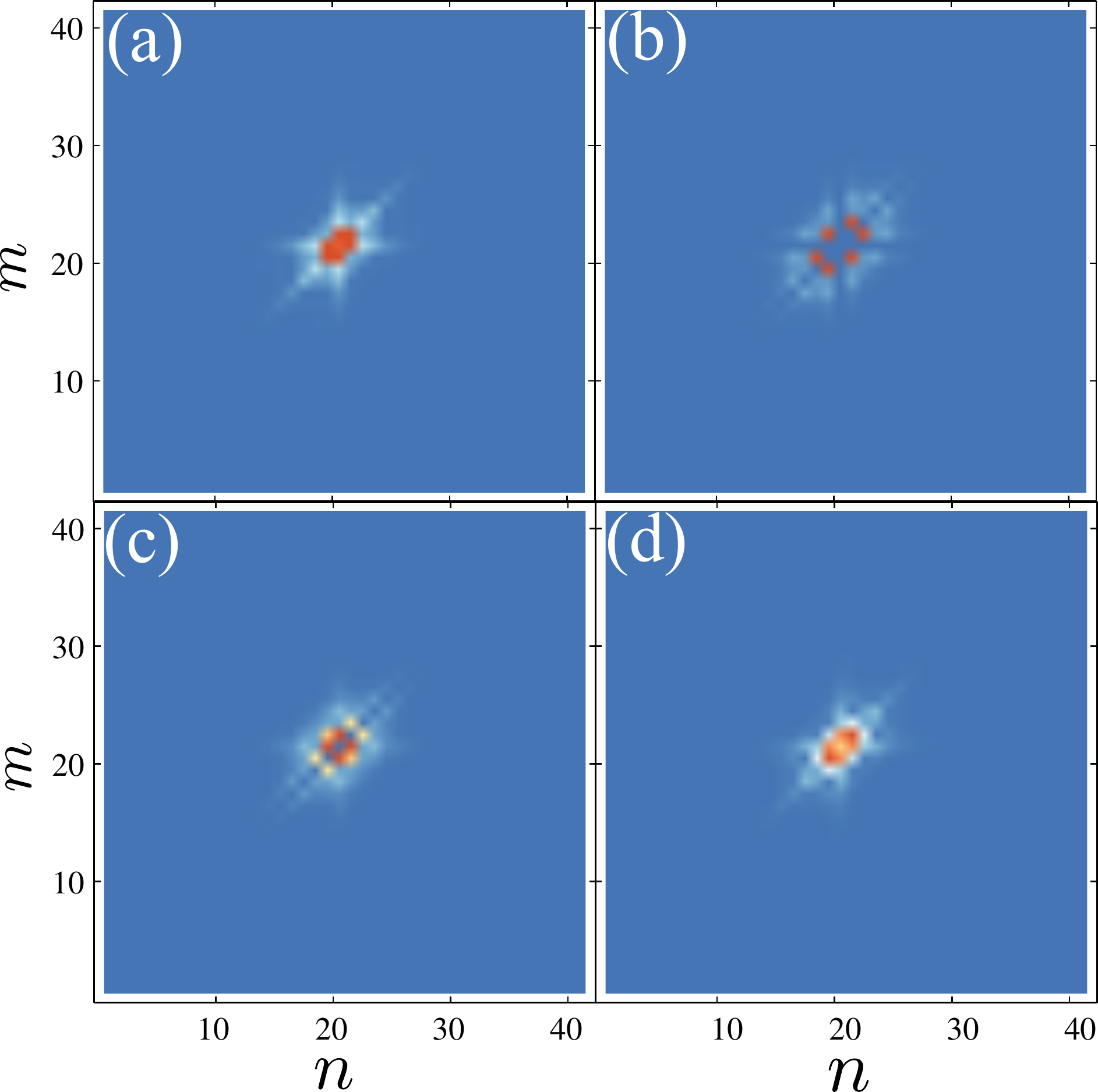}
\caption{The triplon wavefunctions in the coordinate space for $\Omega/J=5$, $\Delta=0$, and $d=1$. (a) The probability to find two QE excitations around one QE excitation at the origin ($f_b$); (b) The probability to find two QE excitations around one photon at the origin ($f_{bba}$); (c) The probability to find two photons around one QE excitation at the origin ($f_{aab}$); (d) The probability to find two photons around one photon at the origin ($f_{a}$).}
\label{W3B}
\end{figure}

In Figs.~\ref{W3B}(a-d), we plot the triplon wavefunctions in real coordinate space, by Fourier transforming the functions $f_{b}(\mathbf{k}_{1},\mathbf{k}_{2})$, $f_{bba}(\mathbf{k}_{1},\mathbf{k}_{2},\mathbf{K})$,  $f_{baa}(\mathbf{k}_{1},\mathbf{k}_{2},\mathbf{K}_{2},\mathbf{K}_{3})$, $f_{a}(\mathbf{k}_{1},\mathbf{k}_{2},\mathbf{K}_{1},\mathbf{K}_{2},\mathbf{K}_{3})$, respectively. As we observe in the figure, the wavefunction has a bound-state behaviour, where the QE excitations and photons are localized around each other. The effective hopping model
\begin{equation}
H_{\rm{triplon}} = \sum_{\mathbf{n,m}}t_{\mathbf{n}-\mathbf{m}}^{T} T_{\mathbf{n}}^{\dagger} T_{\mathbf{m}}
\end{equation}
for the triplon has the same form as Eq.~(\ref{Hd}). In the single triplon subspace, $T_{\mathbf{n}}=|0\rangle\langle\Psi_{3B}(\mathbf{n})|$ is defined by the three-excitation bound state $|\Psi_{3B}(\mathbf{n})\rangle=\sum_{\mathbf{q}}e^{-i\mathbf{q{\cdot}n}}|\Psi_{3B}(\mathbf{q})\rangle/\sqrt{N_{b}}$ localized around QE at the position $\mathbf{n}$, and the effective hopping strength is
\begin{equation}
t_{\mathbf{n}-\mathbf{m}}^{T} = \frac{1}{N_{b}}\sum_{\mathbf{q}}E_{3B}(\mathbf{q}) e^{i\mathbf{q}{\cdot}(\mathbf{n-m})}.
\end{equation}

\subsection{Superfluid to Mott insulator transition}

Let us finally consider the situation of many excitations in the QE array situation. With the intuition we developed with the results of the previous Sections, we know that when the coupling is very strong, $\Omega\gg J$, or when we are in the deep Markov regime, $|\Delta|\gg J,\Omega$, we expect to have very localized bound states around the QEs. However, as one deviates from that conditions, the localization length of the bound states grows leading to strong hybridizing effects between the localized excitations. In this Section, we explore whether this localization length change leads to a superfluid-Mott insulating phase transition in the ground state of the system.

To obtain a detailed quantitative understanding, we study the system numerically using the DMRG method \cite{white92a,schollwock11a}. The DMRG algorithm is a variational method within the class of matrix product states and its nature imposes two constraints on the numerical studies. First, one should adopt open boundary conditions as this is more suitable for DMRG. The Bloch bands and momentum introduced for periodic systems cannot be defined for open systems but the physical properties should be the same if the system is sufficiently large. Secondly, the number of excitations on the bath sites should have an upper bound, that we denote as $\mathcal{C}$, because the computational cost of DMRG is related to the Hilbert space dimensions of the lattice sites. $\mathcal{C}$ should be large enough such that the numerical results reflect the true physics. The ground states have been computed using DMRG in various cases and we find that $\mathcal{C}=5$ is sufficient because increasing it to $6$ does not change the results significantly.

\begin{figure}[tbp]
\centering
\includegraphics[width=0.99\textwidth]{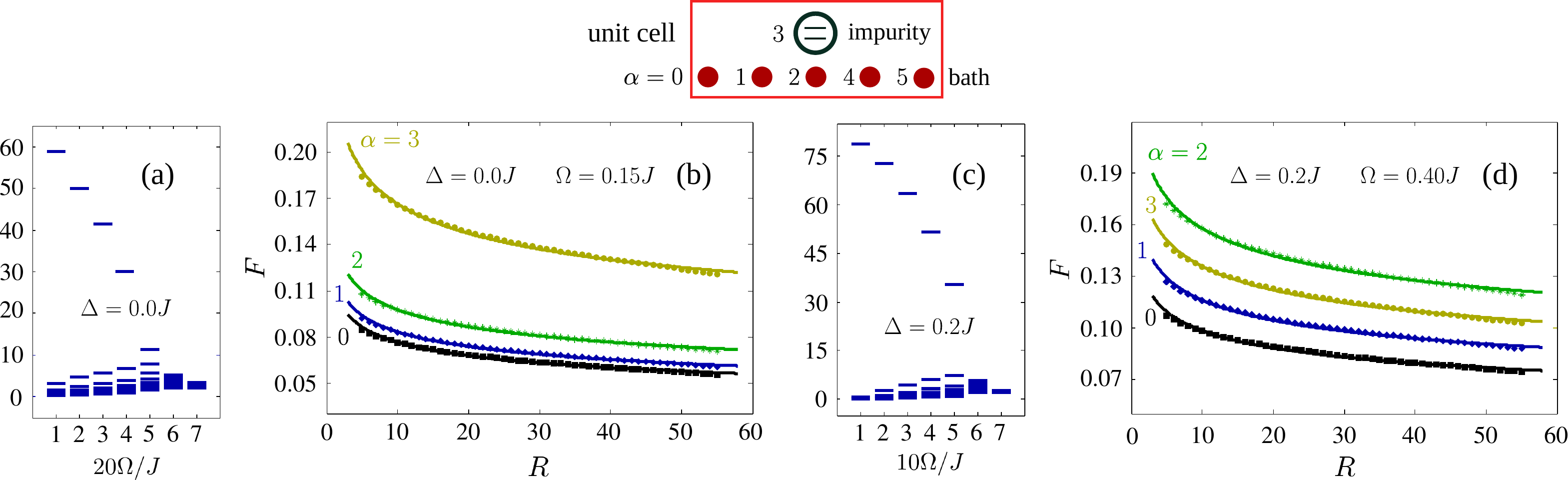}
\caption{The correlation function matrix $F_{i\alpha,j\beta}$ in the system with $N_{\rm{imp}}=N_{\rm{exc}}=80$. The structure of a unit cell is shown on the top. Panels (a) and (c) show the $10$ largest eigenvalues of $F_{i\alpha,j\beta}$. Panels (c) and (d) show the scaling of $F_{i\alpha,j\alpha}$ versus $R=|i-j|$, where the markers are numerical values and the lines are least square fitting results.}
\label{FigureME1}
\end{figure}

\begin{figure}[tbp]
\centering
\includegraphics[width=0.95\textwidth]{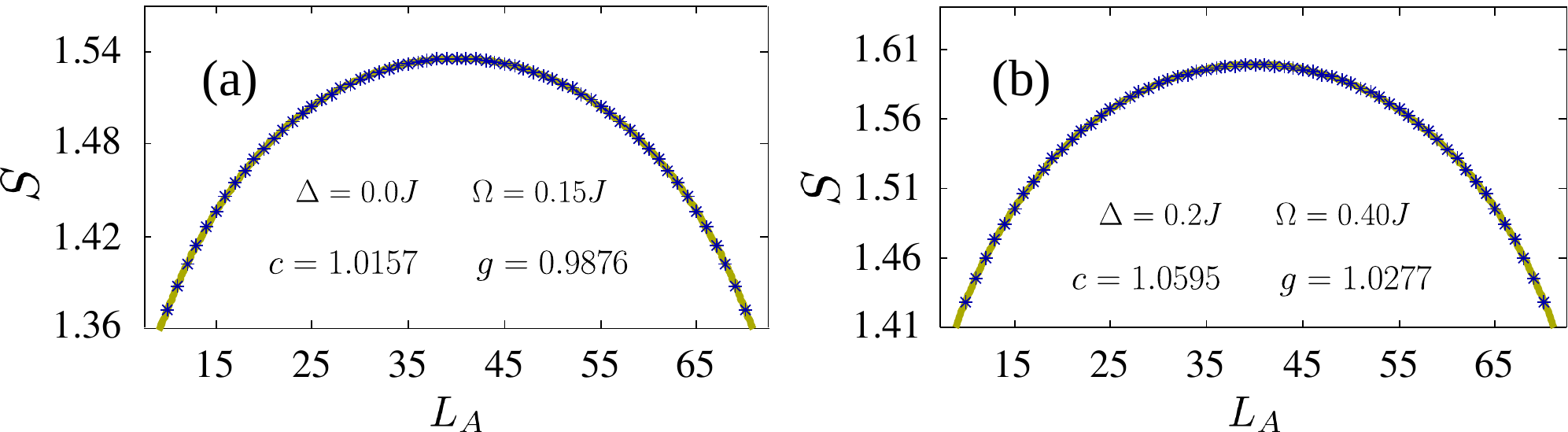}
\caption{The von Neumann entanglement entropy $S$ in the system with $N_{\rm{imp}}=N_{\rm{exc}}=80$. The numerical values are shown as blue stars and the yellow lines are least square fitting results.}
\label{FigureME2}
\end{figure}

The total number of excitations $N_{\rm{exc}}$ is a conserved quantity so different $N_{\rm{exc}}$ sectors can be studied separately. It is possible to access both the superfluid and the Mott insulating phases only if $N_{\rm{exc}}$ is equal to the number of impurity sites $N_{\rm{imp}}$. We label the unit cells using Roman letters $i,j$ etc. and the lattice sites (both the impurity and bath) within a unit cell using Greek letters $\alpha,\beta$ etc. The creation/annihilation operators for both all lattice sites can be expressed on a equal footing as $a_{i\alpha}^{\dagger}$/$a_{i\alpha}$. The first thing we would like to confirm is the existence of two phases in the system. The diagnostic tool we use is the correlation function matrix $F_{i\alpha,j\beta}=\langle a_{i\alpha}^{\dagger}a_{j\beta}\rangle$ with $i\alpha$ ($j\beta$) interpreted as the row (column) index. Fig.~\ref{FigureME1} shows the eigenvalues of $F_{i\alpha,j\beta}$ in the system with $N_{\rm{imp}}=N_{\rm{exc}}=80$ at $\Delta=0.0,0.2$ and various different $\Omega$. The single-excitation bound states in the large $\Omega $ regime are very localized so we expect to see Mott insulator behaviour, where $F_{i\alpha,j\beta}$ has multiple eigenvalues of similar magnitudes corresponding to the multiple modes occupied by the excitations. When the spatial extent of the single-excitation bound states increases, the system transits to the superfluid phase where most excitations occupy the same mode so $F_{i\alpha,j\beta}$ has only one dominant eigenvalue. The low-energy effective theory for the superfluid phase is a Luttinger liquid theory, which predicts that $F_{i\alpha,j\alpha}\sim{|i-j|^{f}}$ to the first order~\cite{haldane81a}. Fig.~\ref{FigureME1} shows two examples of least square fitting of $F_{i\alpha,j\alpha}$, where we choose $i=10$ and $5{\leq}|i-j|{\leq}55$ because the power law scaling is not expected to be accurate if $i$ and/or $j$ are too close to the edge or if $|i-j|$ is too small. One can see that $f$ is basically independent of $\alpha $, with an approximate value of $-0.177$ at $\Delta=0.0J,\Omega=0.15J$ and $-0.154$ at $\Delta=0.2J,\Omega=0.40J$. The von Neumann entanglement entropy also provides valuable information about the system. This quantity is defined as $S=-\rm{Tr}(\rho_{A}\ln\rho_{A})$ where $\rho_{A}$ is the reduced density matrix of the left $L_{A}$ unit cells of the chain. For the superfluid phase, the functional form of $S$ is
\begin{equation}
S(L_{A}) =\frac{c}{6}\ln \left[ \frac{L}{\pi}\sin\left(\pi\frac{L_{A}}{L}\right)\right] + g + F,
\label{EntropyFunction}
\end{equation}
where $c$ is the central charge of the Luttinger liquid, $g$ is a constant, and $F$ is a non-universal oscillating term~\cite{calabrese04a}. Fig.~\ref{FigureME2} shows two examples of least square fitting of $S$ in the system with $N_{\rm{imp}}=N_{\rm{exc}}=80$, where we choose $10{\leq}L_{A}{\leq}70$ and discard those close to $0$ or $L$ to avoid edge effect. It turns out that the oscillating term is negligible and $c=1.0157$ ($c=1.0595$) for $\Delta=0.0J,\Omega=0.15J$ ($\Delta=0.2J,\Omega=0.40J$). This suggests that the superfluid state is a one-component Luttinger liquid with $c=1$. In the Mott insulating phase at larger $\Omega$, $S$ is almost constant in the bulk of the system as one expects for a 1D gapped phase.

\section{Conclusions \label{sec:conclu}}

To sum up, we have studied the emergent dynamics of many QEs interacting with structured photonic reservoirs in the non-Markovian and many excitation regimes. In the two- and three-excitation subspaces, we provide analytical formulas for both the energies and wavefunctions of the relevant states governing the dynamics for arbitrary bath dimension and energy dispersion. We apply these formulas to study the case of a nearest-neighbour tight-binding one-dimensional bath uncovering several phenomena which are oblivious in perturbative descriptions. First, we show the emergence of effective hopping models in parameter regimes far from the Markovian ones, with the advantage of having stronger dipole-dipole couplings as compared to the perturbative regimes. Second, we also predict the emergence of new hopping models in the excited part of the spectrum, in which doublon/triplon states hop between the different QEs coupled to the bath. Finally, we numerically characterize the ground state in the many excitation sector of these quantum optical models, and show how the ground state can undergo an optically driven Mott-superfluid phase transition controlled by the localization length of the bound states. An interesting research direction is to apply the theoretical toolbox developed in the manuscript to study higher dimensional structured baths~\cite{gonzaleztudela17a,gonzaleztudela18c}.

\newpage

\appendix

\section{Single excitation \label{app:1ES}}

In this Appendix, we solve the Schr\"{o}dinger equation to study the single excitation bound states of two-QEs with distance $d=|\mathbf{d}|$ and the QE array with the lattice spacing $z=N/N_{b}$, where $\mathbf{d}=\mathbf{n}_{2}-\mathbf{n}_{1}$ is the vector connecting two quantum emitters (QEs). Without loss of generality, we choose $\mathbf{n}_{1}=0$ and $\mathbf{n}_{2}=\mathbf{d}$ for two QEs. The parameters $u_{1,\lambda}$, $u_{2,\lambda}$, and the wavefunction $f_{\lambda}(\mathbf{k})$ are determined by the Schr\"{o}dinger equation as
\begin{eqnarray}
&& \Delta u_{1,\lambda}+\frac{\Omega}{\sqrt{N}}\sum_{\mathbf{k}} f_{\lambda}(\mathbf{k}) = E_{1\lambda} u_{1,\lambda}, \nonumber \\
&& \Delta u_{2,\lambda}+\frac{\Omega}{\sqrt{N}}\sum_{\mathbf{k}} e^{i\mathbf{k}\cdot\mathbf{d}}f_{\lambda}(\mathbf{k}) = E_{1\lambda}u_{2,\lambda}, \nonumber \\
&& \varepsilon_{\mathbf{k}}f_{\lambda}(\mathbf{k}) + \frac{\Omega}{\sqrt{N}}(u_{1,\lambda}+e^{-i\mathbf{k}\cdot\mathbf{d}} u_{2,\lambda}) = E_{1\lambda}f_{\lambda}(\mathbf{k}).
\label{A1}
\end{eqnarray}

By solving Eq. (\ref{A1}), we obtain the following equation to determine the bound state energies:
\begin{equation}
G^{-1}(E_{1\pm})\mathbf{u}_{\pm}=0\,, \label{SE}
\end{equation}
and the vectors $\mathbf{u}_{\pm}=(u_{1,\pm},u_{2,\pm})^{T}$, where the Green function
\begin{equation}
G(\omega)=\frac{1}{\omega-\Delta-\Sigma_{\rm{d}}(\omega)-\Sigma_{\rm{o}}(\omega)\sigma_{x}}
\end{equation}
is defined by the Pauli matrix $\sigma_{x}$ and the self-energies
\begin{eqnarray}
\Sigma_{\rm{d}}(\omega) &=& \frac{\Omega^{2}}{N} \sum_{\mathbf{k}}\frac{1}{\omega-\varepsilon_{\mathbf{k}}+i0^{+}}, \nonumber \\
\Sigma_{\rm{o}}(\omega) &=& \frac{\Omega^{2}}{N} \sum_{\mathbf{k}}\frac{e^{i\mathbf{k}\cdot\mathbf{d}}}{\omega-\varepsilon_{\mathbf{k}}+i0^{+}}.
\end{eqnarray}

For the (anti-) symmetric bound state with energy $E_{1\pm}$, the parameters $u_{1,\pm}={\pm}u_{2,\pm}{\equiv}u_{\pm}/\sqrt{2}$, and the wavefunction
\begin{equation}
f_{\lambda}(\mathbf{k}) = \frac{\Omega}{\sqrt{2N}} u_{\pm} \frac{1{\pm}e^{-i\mathbf{k}\cdot\mathbf{d}}}{E_{1\pm}-\varepsilon_{\mathbf{k}}},
\end{equation}
where
\begin{equation}
u_{\pm}^{-2}=1+\frac{\Omega^{2}}{N}\sum_{\mathbf{k}}\frac{1{\pm}\cos\mathbf{k}\cdot\mathbf{d}}{(E_{1\pm }-\varepsilon_{\mathbf{k}})^{2}}
\end{equation}
is determined by the normalization condition.

The bound state energies $E_{1\pm}$ and the parameters $u_{\pm}^{2}$ are the poles and the corresponding residues of the Green function
\begin{eqnarray}
G_{\pm}(\omega) &=& \frac{1}{\omega-\Delta-\Sigma_{\rm{d}}(\omega)\mp\Sigma_{\rm{o}}(\omega)} =\sum_{\lambda}\frac{Z_{1\lambda}^{\pm}}{\omega-E_{1\lambda}+i0^{+}},
\end{eqnarray}
where $Z_{1\lambda}^{\pm}$ gives the probability $|\langle{0}|b_{\pm}\beta_{\lambda}^{\dagger}|{0}\rangle|^{2}$ to detect the modes $b_{\pm}^{\dagger}|0\rangle=(b_{1}^{\dagger}{\pm}b_{2}^{\dagger})/\sqrt{2}$ in the eigenstate $\beta_{\lambda}^{\dagger}|0\rangle$. As we show in the main text, there are certain parameter regimes in which the antisymmetric state merges into the continuum and only a single bound state exists. However, when both bound states exist a low-energy effective Hamiltonian can be written:
\begin{equation}
H_{\rm{eff}}^{(1)}=E_{1+}\beta_{+}^{\dagger}\beta_{+}+E_{1-}\beta_{-}^{\dagger}\beta_{-}\,,
\end{equation}
obtained by projecting onto the\ subspace with the symmetric and anti-symmetric bound states.

In the Markovian limit $|\Delta|\gg\Omega$ ($\Delta<0$), the single excitation bound state is extremely localized, such that the effective hopping $|t_{\rm{eff}}|\ll|\Delta|$. Thus, the single particle bound state energies $E_{1\pm}{\sim}E_{1B}{\pm}t_{\rm{eff}}$ can be expanded around the single-excitation bound state energy
\begin{equation}
E_{1B}=\Delta+\frac{\Omega^{2}}{N}\sum_{\mathbf{k}}\frac{1}{\Delta-\varepsilon_{\mathbf{k}}}\sim\Delta  \label{e1B}
\end{equation}
in the presence of one QE. The secular equation determines the hopping strength
\begin{equation}
t_{\rm{eff}}=\frac{\Omega^{2}Z_{1B}}{N}\sum_{\mathbf{k}}\frac{e^{i\mathbf{k\cdot d}}}{E_{1B}-\varepsilon_{\mathbf{k}}}, 
\label{teff}
\end{equation}
where
\begin{equation}
Z_{1B}= \left[ 1+\frac{\Omega^{2}}{N}\sum_{\mathbf{k}}\frac{1}{(E_{1B}-\varepsilon_{\mathbf{k}})^{2}} \right]^{-1}. 
\label{z0}
\end{equation}
In the strong coupling limit $\Omega/J{\gg}1$, the single excitation bound state is also localized, as a result, the effective hopping $t_{\rm{eff}}$ is determined by Eq. (\ref{teff}), where $E_{1B}\sim \Omega$.

For the periodic QE array coupled to the photonic bath, the eigenstates form polariton bands. For the quasi-momentum $\mathbf{p}$, the eigenstate $\beta_{\mathbf{p}\lambda}^{\dagger}|0\rangle$ of Hamiltonian $H_{\mathbf{p}}$ has the energy $E_{1\lambda}(\mathbf{p})$, where the creation operator $\beta_{\mathbf{p}\lambda}^{\dagger}=u_{\mathbf{p}\lambda}b_{\mathbf{p}}^{\dagger}+\sum_{\mathbf{K}}f_{\mathbf{k}\lambda}a_{\mathbf{k}}^{\dagger}$ of the polariton in the $\lambda$ band is determined by the secular equations
\begin{eqnarray}
&& \Delta u_{\mathbf{p}\lambda} + \frac{\Omega}{\sqrt{z}} \sum_{\mathbf{K}} f_{\mathbf{k}\lambda} = E_{1\lambda}(\mathbf{p}) u_{\mathbf{p}\lambda}, \nonumber \\
&& \varepsilon_{\mathbf{k}}f_{\mathbf{k}\lambda}+\frac{\Omega}{\sqrt{z}}u_{\mathbf{p}\lambda} = E_{1\lambda}(\mathbf{p}) f_{\mathbf{k}\lambda}.
\end{eqnarray}

\section{Green functions of Two-excitation in two QEs \label{app:2ES1}}

In this Appendix, we derive the exact form of the two-excitation Green function $G_{2}(\omega)$ for two QEs. The quadratic Hamiltonian in Eq.~(\ref{eq:Ham}) is taken as the unperturbed part. In the interaction picture, the two-particle Green function reads
\begin{equation}
G_{2}(t)=\frac{-i\langle{0}|\mathcal{T}\alpha_{2,I}(t)\alpha_{1,I}(t)\alpha_{1}^{\dagger}\alpha_{2}^{\dagger}e^{-i\int_{-\infty}^{+\infty}H_{I}(t^{\prime})dt^{\prime}}|0\rangle}{\langle{0}|\mathcal{T}e^{-i\int_{-\infty}^{+\infty}H_{I}(t^{\prime})dt^{\prime}}|0\rangle},
\label{G2t}
\end{equation}
where $\alpha_{j,I}(t)$ and $H_{I}(t)$ are the operator $\alpha_{j=1,2}$ and the hard-core interaction $H_{I}=U\sum_{j}b_{j}^{\dagger}b_{j}^{\dagger}b_{j}b_{j}/2$ in the interaction picture.

By expanding the unitary evolution operator in Eq.~(\ref{G2t}), the Fourier transform $G_{2}(\omega)=[G_{2}(\omega)]_{0}+[G_{2}(\omega)]_{c}$ can be written as the free propagation part
\begin{equation}
[G_{2}(\omega)]_{0} = -i\int dte^{i{\omega}t}\langle{0}|\mathcal{T}\alpha_{2,I}(t)\alpha_{1,I}(t)\alpha_{1}^{\dagger}\alpha_{2}^{\dagger}|0\rangle,
\end{equation}
and the connected part
\begin{eqnarray}
[G_{2}(\omega)]_{c} &=& \int_{-\infty}^{+\infty} dt e^{i{\omega}t} \Bigg[ (-i)^{2}\int_{-\infty}^{+\infty} dt_{1} \langle\mathcal{T}\alpha_{2,I}(t)\alpha_{1,I}(t)\alpha_{1}^{\dagger}\alpha_{2}^{\dagger}H_{I}(t_{1})\rangle_{c} \nonumber \\
&\phantom{=}& + \frac{1}{2!}(-i)^{3} \int_{-\infty}^{+\infty} dt_{1}dt_{2} \langle\mathcal{T}\alpha_{2,I}(t)\alpha_{1,I}(t)\alpha_{1}^{\dagger}\alpha_{2}^{\dagger}H_{I}(t_{1})H_{I}(t_{2})\rangle_{c} \nonumber \\
&\phantom{=}& + \cdots \Bigg],
\end{eqnarray}
where $\langle\cdots\rangle_{c}$ denotes the connected Green function on the vacuum state.

Using the Wick theorem, we obtain
\begin{equation}
[G_{2}(\omega)]_{c} = 2 \sum_{jj^{\prime}} \Pi_{\alpha_{1}\alpha_{2}}^{j}(\omega) T_{jj^{\prime}}(\omega)\bar{\Pi}_{\alpha_{1}\alpha_{2}}^{j^{\prime}}(\omega)
\end{equation}
by the convolutions
\begin{eqnarray}
\Pi_{\alpha_{1}\alpha_{2}}^{j}(\omega) &=& i\int\frac{d\omega^{\prime}}{2\pi} G_{\alpha_{1}j}(\omega^{\prime}) G_{\alpha_{2}j}(\omega-\omega^{\prime}), \nonumber \\
\bar{\Pi}_{\alpha_{1}\alpha_{2}}^{j}(\omega) &=& i\int\frac{d\omega^{\prime}}{2\pi} G_{j\alpha_{1}}(\omega^{\prime}) G_{j\alpha_{2}}(\omega-\omega^{\prime}), \nonumber \\
\Pi_{jj^{\prime}}(\omega) &=& i\int\frac{d\omega^{\prime}}{2\pi} G_{jj^{\prime}}(\omega^{\prime}) G_{jj^{\prime}}(\omega-\omega^{\prime}),
\end{eqnarray}
and the Dyson expansion
\begin{eqnarray}
T_{jj^{\prime}}(\omega) = U\delta_{jj^{\prime}} + U\Pi_{jj^{\prime}}(\omega)U + \cdots = U\delta_{jj^{\prime}} + \sum_{j_{1}} U \Pi_{jj_{1}}(\omega) T_{j_{1}j^{\prime}}(\omega)
\label{T2t}
\end{eqnarray}
of the scattering $T$-matrix, where $G_{{\alpha}j}(\omega) = \int dt G_{{\alpha}j}(t) e^{i{\omega}t}$, $G_{j\alpha}(\omega) = \int dt G_{j\alpha}(t) e^{i{\omega}t}$, and $G_{jj^{\prime}}(\omega) = \int dt G_{jj^{\prime}}(t) e^{i{\omega}t}$ are the Fourier transforms of the single-excitation Green functions $G_{{\alpha}j}(t)=-i\langle{0}|\alpha(t)b_{j}^{\dagger}|0\rangle\theta(t)$, $G_{j\alpha}(t)=-i\langle{0}|b_{j}(t)\alpha^{\dagger}|0\rangle\theta(t)$, and $G_{jj^{\prime}}(t)=-i\langle{0}|b_{j}(t)b_{j^{\prime}}^{\dagger}|0\rangle\theta(t)$, respectively. Solving the matrix Eq.~(\ref{T2t}), we obtain $T(\omega)=-\Pi^{-1}(\omega)$.

Due to the fact $G_{11}(\omega)=G_{22}(\omega)$ and $G_{12}(\omega)=G_{21}(\omega)$, the $T$-matrix can be diagonalized as
\begin{equation}
T(\omega)=\sum_{s=\pm}T_{s}(\omega)|s\rangle\langle{s}|
\end{equation}
in the symmetric and anti-symmetric scattering channels $|\pm\rangle=(1,{\pm}1)/\sqrt{2}$, where the eigenvalues are
\begin{eqnarray}
T_{s=\pm}(\omega) = -\frac{1}{\Pi_{11}(\omega)+s\Pi_{12}(\omega)} = -\left( \frac{1}{2}\sum_{\lambda\lambda^{\prime}, \sigma=\pm} \frac{Z_{1\lambda}^{\sigma} Z_{1\lambda^{\prime}}^{s\sigma}}{\omega-E_{1\lambda}-E_{1\lambda^{\prime}}} \right)^{-1}.
\end{eqnarray}
The connected Green function in the diagonalized basis becomes
\begin{equation}
[G_{2}(\omega)]_{c} = \sum_{s=\pm} \Pi_{\alpha_{1}\alpha_{2}}^{s}(\omega)T_{s}(\omega)\bar{\Pi}_{\alpha_{1}\alpha_{2}}^{s}(\omega),
\end{equation}
where the ``bubble" term
\begin{equation}
\Pi_{\alpha_{1}\alpha_{2}}^{s}(\omega) = \sum_{j=1,2}s^{j-1} \Pi_{\alpha_{1}\alpha_{2}}^{j}(\omega),\bar{\Pi}_{\alpha_{1}\alpha_{2}}^{s}(\omega) = \sum_{j=1,2}s^{j-1} \bar{\Pi}_{\alpha_{1}\alpha_{2}}^{j}(\omega).
\label{PI}
\end{equation}

In the interacting channel $s=\pm $, the bound state energy $E_{2s}$ is determined by the pole of $T_{s}(\omega )$, i.e., $T_{s}(E_{2s})=0$. The corresponding residue
\begin{equation}
Z_{0}^{s} = \left( \frac{1}{2}\sum_{\lambda\lambda^{\prime},\sigma=\pm}Z_{1\lambda}^{\sigma}Z_{1\lambda^{\prime}}^{s\sigma}h_{s,\lambda\lambda^{\prime}}^{-2} \right)^{-1}
\end{equation}
in the vicinity of the pole $E_{2s}$ and $\Pi_{\alpha_{1}\alpha_{2}}^{s}(E_{2s})$ result in the probability
\begin{equation}
Z_{2s} = \frac{Z_{0}^{s}}{4} \left( \sum_{\lambda\lambda^{\prime},\sigma=\pm}\sigma Z_{1\lambda}^{\sigma}Z_{1\lambda^{\prime}}^{s\sigma} h_{s,\lambda\lambda^{\prime}}^{-1} \right)^{2}
\label{Z2s}
\end{equation}
to detect two QEs in the excited states, the wavefunction
\begin{equation}
\varphi_{js}(\mathbf{k})=\sqrt{\frac{Z_{0}^{s}}{N}}\sum_{\lambda\lambda^{\prime},\sigma=\pm}\frac{\Omega Z_{1\lambda}^{\sigma} Z_{1\lambda^{\prime}}^{s\sigma}\sigma^{j-1}}{2h_{s,\lambda \lambda ^{\prime }}h_{s,\mathbf{k}\lambda}} \left( 1+s{\sigma}e^{-i\mathbf{k}\cdot\mathbf{d}} \right)
\label{ph1}
\end{equation}
of the $j$-QE in the excited state with one photon of momentum $\mathbf{k}$ in the bath, and the amplitude
\begin{eqnarray}
\varphi_{2s}(\mathbf{k},\mathbf{k}^{\prime}) &=& \frac{\sqrt{Z_{0}^{s}}}{4N} \sum_{\lambda\lambda^{\prime},\sigma=\pm} \frac{ \Omega^{2} Z_{1\lambda}^{\sigma} Z_{1\lambda^{\prime}}^{s\sigma} }{ h_{s,\mathbf{k}\lambda} h_{s,\mathbf{k}^{\prime}\lambda^{\prime}} } \left( h_{s,\lambda\lambda^{\prime}}^{-1} + h_{s,\mathbf{kk}^{\prime}}^{-1} \right) \nonumber \\
&\phantom{=}& \times \left( 1+s{\sigma}e^{-i\mathbf{k}\cdot\mathbf{d}} \right) \left( 1+{\sigma}e^{-i\mathbf{k}^{\prime}\cdot\mathbf{d}} \right)
\label{ph2}
\end{eqnarray}
to find two photons with momenta $\mathbf{k}$ and $\mathbf{k}^{\prime}$, where $h_{s,\lambda\lambda^{\prime}}=E_{2s}-E_{1\lambda }-E_{1\lambda^{\prime}}$, $h_{s,\mathbf{k}\lambda}=E_{2s}-\varepsilon_{\mathbf{k}}-E_{1\lambda}$, and $h_{s,\mathbf{kk}^{\prime}}=E_{2s}-\varepsilon_{\mathbf{k}}-\varepsilon_{\mathbf{k}^{\prime}}$.

The ground state analyzed in Sec.~\ref{sec:2EGS:1} has the smallest energy in the symmetric subspace $s=+$. The symmetric and anti-symmetric doublon states studied in Sec.~\ref{sec:2EDL:1} correspond to the isolated pole with higher energy in the subspace $s=+$ and that in the subspace $s=-$. The eigenenergies and the wavefunctions of the ground state and the doublon states are determined by $T_{s}(E_{2s})=0$ and Eqs.~(\ref{Z2s})-(\ref{ph2}).

The doublon states with higher energies in the $s=\pm$ channels only exist for certain parameters. The symmetric bound state exists if $E_{1+}>2E_{1-}$ and $T_{+}(E_{1+})>0$, while the anti-symmetric bound state can be found if the single-excitation bound state $\beta_{-}^{\dagger}|0\rangle$ exists and $T_{-}(E_{1+})>0$. In the Markovian limit and strong coupling regime, one can always find two higher-energy bound states in the $\pm $ channels.

\section{Green functions of two-excitation in QE array \label{app:2ES2}}

In this Appendix, we derive the equation to determine the band structure in the two-excitation spectrum, and the two-excitation wavefunctions $f_{b}(\mathbf{p})$, $f_{ba}(\mathbf{p,K})$, and $f_{a}(\mathbf{p},\mathbf{K,K}^{\prime})$.

The band structure can be identified by the position of the poles and branch cuts of $G_{2}(\mathbf{q},\omega )$ for two excitations $\alpha_{1}^{\dagger}\alpha_{2}^{\dagger}|0\rangle$ with momenta $\mathbf{p}_{\pm}=\mathbf{q}/2{\pm}\mathbf{p}$. In the interaction picture, the Green function $G_{2}(\mathbf{q},t)$ in the time domain reads
\begin{equation}
G_{2}(\mathbf{q},t) = -i\frac{\langle{0}|\mathcal{T}\alpha_{2,I}(t)\alpha_{1,I}(t)\alpha_{1}^{\dagger}\alpha_{2}^{\dagger} e^{-i\int_{-\infty}^{+\infty} H_{\rm{hc}}(t^{\prime}) dt^{\prime}} |0\rangle}{\langle{0}|\mathcal{T}e^{-i\int_{-\infty}^{+\infty}H_{I}(t^{\prime}) dt^{\prime}}|0\rangle}.
\end{equation}
The expansion of the unitary evolution operator gives rise to the Fourier transform $G_{2}(\mathbf{q},\omega)=[G_{2}(\mathbf{q},\omega)]_{0}+[G_{2}(\mathbf{q},\omega)]_{c}$, where
\begin{equation}
[G_{2}(\mathbf{q},\omega)]_{0} = -i\int dt e^{i{\omega}t}\langle{0}|\alpha_{2,I}(t)\alpha_{1,I}(t)\alpha _{1}^{\dagger}\alpha_{2}^{\dagger}|0\rangle
\end{equation}
describes the free propagation part, and
\begin{eqnarray}
[G_{2}(\mathbf{q},\omega)]_{c} &=& \int dt e^{i{\omega}t} \Bigg[ (-i)^{2} \int dt_{1} \langle\mathcal{T}\alpha_{2,I}(t)\alpha_{1,I}(t)\alpha_{1}^{\dagger}\alpha_{2}^{\dagger} H_{\rm{hc}}(t_{1}) \rangle_{c} \nonumber \\
&\phantom{=}& + \frac{1}{2}(-i)^{3} \int dt_{1} dt_{2} \langle \mathcal{T}\alpha_{2,I}(t)\alpha_{1,I}(t)\alpha_{1}^{\dagger}\alpha_{2}^{\dagger} H_{\rm{hc}}(t_{1}) H_{\rm{hc}}(t_{2}) \rangle_{c} \nonumber \\
&\phantom{=}&  + \cdots \Bigg]
\end{eqnarray}
is the connected part.

Using Wick theorem, we obtain
\begin{equation}
[G_{2}(\mathbf{q},\omega)]_{c} = \frac{2}{N_{b}} \Pi_{\alpha_{1}\alpha_{2}}(\mathbf{q},\mathbf{p},\omega ) T(\mathbf{q},\omega) \bar{\Pi}_{\alpha_{1}\alpha_{2}}(\mathbf{q},\mathbf{p},\omega)
\end{equation}
by the convolutions
\begin{eqnarray}
\Pi_{\alpha_{1}\alpha_{2}}(\mathbf{q},\mathbf{p},\omega) &=& i\int\frac{d\omega^{\prime}}{2\pi} G_{\alpha_{1}b_{\mathbf{p}_{+}}}(\omega^{\prime}) G_{\alpha_{1}b_{\mathbf{p}_{-}}}(\omega-\omega^{\prime}), \nonumber \\
\bar{\Pi}_{\alpha_{1}\alpha_{2}}(\mathbf{q},\mathbf{p},\omega) &=& i\int\frac{d\omega^{\prime}}{2\pi} G_{b_{\mathbf{p}_{+}}\alpha_{1}}(\omega^{\prime}) G_{b_{\mathbf{p}_{-}}\alpha_{2}}(\omega-\omega^{\prime}),
\nonumber \\
\Pi_{b}(\mathbf{q},\omega) &=& \frac{1}{N_{b}} \sum_{\mathbf{p}^{\prime}} i \int \frac{d\omega^{\prime}}{2\pi} G_{b}(\frac{\mathbf{q}}{2}+\mathbf{p}^{\prime},\omega^{\prime}) G_{b}(\frac{\mathbf{q}}{2}-\mathbf{p}^{\prime},\omega-\omega^{\prime}),
\end{eqnarray}
and the scattering $T$ matrix
\begin{eqnarray}
T(\mathbf{q},\omega) = U + U\Pi_{b}(\mathbf{q},\omega)U + \cdots = U + U\Pi_{b}(\mathbf{q},\omega)T(\mathbf{q},\omega),
\label{T2m}
\end{eqnarray}
where the Green functions are $G_{\alpha_{i}b_{\mathbf{p}}}(\omega) = -i\int dt e^{i{\omega}t} \langle \alpha_{i}(t)b_{\mathbf{p}}^{\dagger} \rangle \theta(t)$, $G_{b_{\mathbf{p}}\alpha_{i}}(\omega)=-i\int
dt e^{i{\omega}t} \langle b_{\mathbf{p}}(t) \alpha_{i}^{\dagger} \rangle \theta(t)$, and $G_{b}(\mathbf{p},\omega) = -i \int dt e^{i{\omega}t} \langle b_{\mathbf{p}}(t) b_{\mathbf{p}}^{\dagger} \rangle \theta (t)$, respectively. Solving Eq.~(\ref{T2m}), we obtain
\begin{eqnarray}
T(\mathbf{q},\omega) &=& - \Pi_{b}^{-1}(\mathbf{q},\omega+i0^{+}) \nonumber \\
                     &=& - \left[ \frac{1}{N_{b}} \sum_{\mathbf{p},\lambda\lambda^{\prime}} \frac{Z_{1\lambda}(\mathbf{p}) Z_{1\lambda^{\prime}}(\mathbf{q}-\mathbf{p})}{\omega-E_{1\lambda}(\mathbf{p})-E_{1\lambda^{\prime}}(\mathbf{q}-\mathbf{p})} \right]^{-1}
\end{eqnarray}
and $[G_{2}(\mathbf{q},\omega)]_{c}$.

The poles and residues of $T_{2}(\mathbf{q},\omega)$ determine the doublon dispersion relation $E_{2B}(\mathbf{q})$ and band structure of scattering states, respectively. For the doublon state $|\Psi_{2D}(\mathbf{q})\rangle$, the residue of $T_{2}(\mathbf{q,}\omega)$ in the vicinity of $E_{2B}(\mathbf{q})$ and $\Pi_{\alpha_{1}\alpha_{2}}(\mathbf{q},\mathbf{p},E_{2B}(\mathbf{q}))$ give rise to the wavefunctions\begin{eqnarray}
f_{b}(\mathbf{p}) = \sum_{\lambda\lambda^{\prime}} \sqrt{ \frac{Z_{2}(\mathbf{q})}{2N_{b}h_{\lambda\lambda^{\prime}}^{2}(\mathbf{p})} } Z_{1\lambda}(\mathbf{p}_{+}) Z_{1\lambda^{\prime}}(\mathbf{p}_{-}), \nonumber \\
f_{ab}(\mathbf{k,K}) = \Omega \sqrt{ \frac{2Z_{2}(\mathbf{q})}{N_{b}z}} \sum_{\lambda\lambda^{\prime}}\frac{Z_{1\lambda}(\mathbf{p}_{+}) Z_{1\lambda^{\prime}}(\mathbf{p}_{-})} {h_{\lambda\lambda^{\prime}}(\mathbf{p}) h_{\lambda\mathbf{K}}(\mathbf{p})}, \nonumber \\
f_{a}(\mathbf{k},\mathbf{K,K}^{\prime}) = \Omega^{2} \sqrt{\frac{Z_{2}(\mathbf{q})}{2N_{b}}} \sum_{\lambda\lambda^{\prime}} Z_{1\lambda}(\mathbf{p}_{+}) Z_{1\lambda^{\prime}}(\mathbf{p}_{-}) \frac{ \left[ h_{\lambda\lambda^{\prime}}^{-1}(\mathbf{p})+h_{\mathbf{KK}^{\prime}}^{-1}(\mathbf{p})\right] }{ z h_{\lambda^{\prime}\mathbf{K}}(-\mathbf{p}) h_{\lambda\mathbf{K}^{\prime}}(\mathbf{p})},
\end{eqnarray}
where $h_{\lambda\lambda^{\prime}}(\mathbf{p})=E_{2B}(\mathbf{q})-E_{1\lambda}(\mathbf{p}_{+})-E_{1\lambda^{\prime}}(\mathbf{p}_{-})$, $h_{\lambda\mathbf{K}}(\mathbf{p})=E_{2B}(\mathbf{q})-E_{1\lambda}(\mathbf{p}_{+})-\varepsilon_{\mathbf{K}+\mathbf{p}_{-}}$, and $h_{\mathbf{KK}^{\prime}}(\mathbf{p})=E_{2B}(\mathbf{q})-\varepsilon_{\mathbf{K}+\mathbf{p}_{+}}-\varepsilon_{\mathbf{K}^{\prime}+\mathbf{p}_{-}}$. The wavefunctions can be Fourier transformed to real space to give $f_{b}(\mathbf{r})$, $f_{ab}(\mathbf{r,m})$, and $f_{a}(\mathbf{r},\mathbf{n,m})$, which represent the amplitudes of having two excited QEs that are separated by a distance $\mathbf{r}$, having one photon at the position $\mathbf{m}$ and one excited QE separated from the photon by a distance $\mathbf{r}$, and having two photons at the positions $\mathbf{n}$ and $\mathbf{m}$.

\section{Three-excitation spectrum \label{app:3ES}}

In this Appendix, we study the properties of three excitations in the QE array by the Green function approach. The triplon state has the form
\begin{eqnarray}
&& |\Psi_{3B}(\mathbf{q})\rangle = \sum_{\mathbf{k}_{1}\mathbf{k}_{2}} f_{b}(\mathbf{k}_{1},\mathbf{k}_{2}) b_{\mathbf{k}_{1}}^{\dagger} b_{\mathbf{k}_{2}}^{\dagger} b_{\mathbf{q}-\mathbf{k}_{1}-\mathbf{k}_{2}}^{\dagger} |0\rangle \nonumber \\
&\phantom{=}& + \sum_{\mathbf{k}_{1}\mathbf{k}_{2}\mathbf{K}} f_{bba}(\mathbf{k}_{1},\mathbf{k}_{2},\mathbf{K}) b_{\mathbf{k}_{1}}^{\dagger} b_{\mathbf{k}_{2}}^{\dagger} a_{\mathbf{q}-\mathbf{k}_{1}-\mathbf{k}_{2},\mathbf{K}}^{\dagger} |0\rangle \nonumber \\
&\phantom{=}& + \sum_{\mathbf{k}_{1}\mathbf{k}_{2}\mathbf{K}_{2}\mathbf{K}_{3}} f_{baa}(\mathbf{k}_{1},\mathbf{k}_{2},\mathbf{K}_{2},\mathbf{K}_{3}) b_{\mathbf{k}_{1}}^{\dagger} a_{\mathbf{k}_{2},\mathbf{K}_{2}}^{\dagger} a_{\mathbf{q}-\mathbf{k}_{1}-\mathbf{k}_{2},\mathbf{K}_{3}}^{\dagger} |0\rangle \nonumber \\
&\phantom{=}& + \sum_{\mathbf{k}_{1}\mathbf{k}_{2}\mathbf{K}_{1}\mathbf{K}_{2}\mathbf{K}_{3}} f_{a}(\mathbf{k}_{1},\mathbf{k}_{2},\mathbf{K}_{1},\mathbf{K}_{2},\mathbf{K}_{3}) a_{\mathbf{k}_{1},\mathbf{K}_{1}}^{\dagger}a_{\mathbf{k}_{2},\mathbf{K}_{2}}^{\dagger}a_{\mathbf{q}-\mathbf{k}_{1}-\mathbf{k}_{2},\mathbf{K}_{3}}^{\dagger} |0\rangle.
\end{eqnarray}

Similar to the analysis of two excitations, we introduce the three-excitation Green function
\begin{equation}
G_{3}(t) = -i\langle{0}|\mathcal{T}\alpha_{3}(t)\alpha_{2}(t)\alpha_{1}(t)\alpha_{1}^{\dagger}\alpha_{2}^{\dagger}\alpha_{3}^{\dagger}|0\rangle,
\end{equation}
whose Fourier transform $G_{3}(\omega)=\int dt e^{i{\omega}t} G_{3}(t)$ determines the three-excitation spectrum including scattering and triplon bands. The three excitations have the total quasi-momentum $\mathbf{q}$. The wavefunctions $f_{b}(\mathbf{k}_{1},\mathbf{k}_{2})$, $f_{bba}(\mathbf{k}_{1},\mathbf{k}_{2},\mathbf{K})$, $f_{baa}(\mathbf{k}_{1},\mathbf{k}_{2},\mathbf{K}_{2},\mathbf{K}_{3})$, and $f_{a}(\mathbf{k}_{1},\mathbf{k}_{2},\mathbf{K}_{1},\mathbf{K}_{2},\mathbf{K}_{3})$ are obtained by the residues of Green functions $G_{3}(\omega)$ in the vicinity of poles with (a) $\alpha_{1}=b_{\mathbf{k}_{1}}$, $\alpha_{2}=b_{\mathbf{k}_{2}}$, $\alpha_{3}=b_{\mathbf{q}-\mathbf{k}_{1}-\mathbf{k}_{2}}$; (b) $\alpha_{1}=b_{\mathbf{k}_{1}}$, $\alpha_{2}=b_{\mathbf{k}_{2}}$, $\alpha_{3}=a_{\mathbf{q}-\mathbf{k}_{1}-\mathbf{k}_{2},\mathbf{K}}$; (c) $\alpha_{1}=b_{\mathbf{k}_{1}}$, $\alpha_{2}=a_{\mathbf{k}_{2},K_{2}}$, $\alpha_{3}=a_{\mathbf{q}-\mathbf{k}_{1}-\mathbf{k}_{2},\mathbf{K}_{3}}$; and (d) $\alpha_{1}=a_{\mathbf{k}_{1},\mathbf{K}_{1}}$, $\alpha_{2}=a_{\mathbf{k}_{2},\mathbf{K}_{2}}$, $\alpha_{3}=a_{\mathbf{q}-\mathbf{k}_{1}-\mathbf{k}_{2},\mathbf{K}_{3}}$.

In the interaction picture, the Green function $G_{3}(t)$ becomes
\begin{equation}
G_{3}(t) = -i\frac{\langle{0}|\mathcal{T}\alpha_{3,I}(t)\alpha_{2,I}(t)\alpha_{1,I}(t)\alpha_{1}^{\dagger}\alpha_{2}^{\dagger}\alpha_{3}^{\dagger}e^{-i\int_{-\infty}^{+\infty} H_{\rm{hc}}(t^{\prime
}) dt^{\prime}} |0\rangle} {\langle{0}|\mathcal{T}e^{-i\int_{-\infty}^{+\infty} H_{\rm{hc}}(t^{\prime}) dt^{\prime}} |0\rangle}.
\end{equation}
Using the Dyson expansion, the connected part $[G_{3}(\omega)]_{c}$ reads
\begin{eqnarray}
[G_{3}(\omega)]_{c} &=& \int \frac{d\omega_{1}}{2\pi} \int \frac{d\omega_{1}^{\prime}}{2\pi} \int \frac{d\omega_{2}}{2\pi} \int \frac{d\omega_{2}^{\prime}}{2\pi} \nonumber \\
&\phantom{=}& \frac{1}{N_{b}} P G_{\alpha_{1}b_{\mathbf{p}_{1}}}(\omega_{1}) T(\mathbf{q}-\mathbf{p}_{1},\omega-\omega_{1}) \nonumber \\
&\phantom{=}& \times G_{\alpha_{2} b_{\mathbf{p}_{2}}}(\omega_{2}) G_{\alpha _{3}b_{\mathbf{p}_{3}}}(\omega-\omega_{1}-\omega_{2}) \nonumber \\
&\phantom{=}& \times T_{3}(\mathbf{q},\mathbf{p}_{1},\mathbf{p}_{1}^{\prime};\omega,\omega_{1},\omega_{1}^{\prime}) \nonumber \\
&\phantom{=}& \times \frac{1}{N_{b}} P G_{b_{\mathbf{p}_{1}^{\prime}} \alpha_{1}}(\omega_{1}^{\prime}) T(\mathbf{q}-\mathbf{p}_{1}^{\prime},\omega-\omega_{1}^{\prime}) \nonumber \\
&\phantom{=}& \times G_{b_{\mathbf{p}_{2}^{\prime}}\alpha_{2}}(\omega_{2}^{\prime}) G_{b_{\mathbf{p}_{3}^{\prime}} \alpha_{3}}(\omega-\omega_{1}^{\prime}-\omega_{2}^{\prime}),
\end{eqnarray}
where $P$ denotes the permutation of $\alpha_{1,2,3}$, and the three-excitation $T$-matrix satisfies the equation
\begin{eqnarray}
&& T_{3}(\mathbf{q},\mathbf{p}_{1},\mathbf{p}_{1}^{\prime};\omega,\omega_{1},\omega_{1}^{\prime}) = G_{b}(\mathbf{q}-\mathbf{p}_{1}-\mathbf{p}_{1}^{\prime},\omega-\omega_{1}-\omega_{1}^{\prime}) \nonumber \\
&& + \frac{2}{N_{b}} \sum_{\mathbf{k}} i\int \frac{d\omega_{\mathbf{k}}}{2\pi} G_{b}(\mathbf{q-\mathbf{p}_{1}-k},\omega-\omega_{1}-\omega_{\mathbf{k}}) \nonumber \\
&& \times G_{b}(\mathbf{k},\omega_{\mathbf{k}}) T(\mathbf{q}-\mathbf{k},\omega-\omega_{\mathbf{k}}) T_{3}(\mathbf{q},\mathbf{k},\mathbf{p}_{1}^{\prime};\omega,\omega_{\mathbf{k}},\omega_{1}^{\prime}).
\label{T3}
\end{eqnarray}

The branch cuts of $T_{3}$ correspond to the scattering bands describing both the scattering of three individual polaritons and that between a single doublon and one polariton. The pole of $T_{3}$ determines the triplon band, where three polaritons form the bound state and co-propagate on the lattice.

In the vicinity of the triplon pole, the three-excitation $T$-matrix has the form
\begin{equation}
T_{3}(\mathbf{q},\mathbf{p}_{1},\mathbf{p}_{1}^{\prime};\omega,\omega_{1},\omega_{1}^{\prime}) = \frac{F(\mathbf{p}_{1},\omega_{1}) \bar{F}(\mathbf{p}_{1}^{\prime},\omega_{1}^{\prime})}{\omega-E_{3\mathbf{q}}+i0^{+}}.
\end{equation}
It follows from Eq. (\ref{T3}) that the residue function $F(\mathbf{p}_{1},\omega_{1})$ obeys
\begin{eqnarray}
F(\mathbf{p}_{1},\omega _{1}) &=& \frac{2}{N_{b}} \sum_{\mathbf{k}} i\int \frac{d\omega_{\mathbf{k}}}{2\pi} G_{b}(\mathbf{q-\mathbf{p}_{1}-k},E_{3B}(\mathbf{q})-\omega_{1}-\omega_{\mathbf{k}}) \nonumber \\
&\phantom{=}& \times T(\mathbf{q}-\mathbf{k},E_{3B}(\mathbf{q})-\omega _{\mathbf{k}}) G_{b}(\mathbf{k},\omega_{\mathbf{k}}) F(\mathbf{k},\omega_{\mathbf{k}}).
\end{eqnarray}
By the residue theorem, one can carry out the integral over $\omega_{\mathbf{k}}$ and obtain
\begin{eqnarray}
F(\mathbf{p}_{1},\omega_{1}) &=& \frac{2}{N_{b}} \sum_{\mathbf{k}\lambda} Z_{1\lambda}(\mathbf{k}) G_{b}(\mathbf{q-\mathbf{p}_{1}-k},E_{3B}(\mathbf{q})-\omega_{1}-E_{1\lambda}(\mathbf{k})) \nonumber \\
&\phantom{=}& \times T(\mathbf{q}-\mathbf{k},E_{3B}(\mathbf{q})-E_{1\lambda}(\mathbf{k})) F(\mathbf{k},E_{1\lambda}(\mathbf{k})).
\label{Fp}
\end{eqnarray}

Setting $\omega_{1}=E_{1\lambda_{1}}(\mathbf{p}_{1})$, we can establish the matrix equation
\begin{equation}
f_{\mathbf{p}\lambda_{1}} = \sum_{\mathbf{k}\lambda}\mathbf{M}_{\mathbf{p}\lambda_{1},\mathbf{k}\lambda}(E_{3B}(\mathbf{q}))f_{\mathbf{k}\lambda}
\end{equation}
for $f_{\mathbf{k}\lambda}=F(\mathbf{k},E_{1\lambda}(\mathbf{k}))$, where the matrix
\begin{eqnarray}
\mathbf{M}_{\mathbf{p}\lambda_{1},\mathbf{k}\lambda}[E_{3B}(\mathbf{q})] &=& \frac{2}{N_{b}} Z_{1\lambda}(\mathbf{k}) T(\mathbf{q}-\mathbf{k},E_{3B}(\mathbf{q})-E_{1\lambda}(\mathbf{k})) \nonumber \\
&\phantom{=}& \times G_{b}(\mathbf{q-\mathbf{p}-k},E_{3B}(\mathbf{q})-E_{1\lambda_{1}}(\mathbf{p})-E_{1\lambda}(\mathbf{k})).
\end{eqnarray}
The triplon energy is determined by $\det[\mathbf{M}(E_{3B}(\mathbf{q}))\mathbf{-I}]=0$, and $f_{\mathbf{k}\lambda }$ is the eigenstate corresponding to the zero eigenvalue of $\mathbf{M}(E_{3B}(\mathbf{q}))\mathbf{-I}$, which gives rise to the residue function $F(\mathbf{p}_{1},\omega_{1})$ by Eq.~(\ref{Fp}).

The residues in the vicinity of the pole $E_{3B}(\mathbf{q})$ leads to the wavefunctions
\begin{eqnarray}
\fl f_{b}(\mathbf{p}_{1},\mathbf{p}_{2}) = \frac{1}{6N_{b}} P \sum_{\lambda_{1}\lambda_{2}} Z_{1\lambda_{1}}(\mathbf{p}_{1}) Z_{1\lambda _{2}}(\mathbf{p}_{2}) T(\mathbf{q}-\mathbf{p}_{1},E_{3B}(\mathbf{q})-E_{1\lambda_{1}}(\mathbf{p}_{1})) \nonumber \\
\fl \times G_{b}(\mathbf{p}_{3},E_{3B}(\mathbf{q})-E_{1\lambda _{1}}(\mathbf{p}_{1})-E_{1\lambda_{2}}(\mathbf{p}_{2})) f_{\mathbf{p}_{1}\lambda_{1}}, \nonumber \\
\fl f_{bba}(\mathbf{p}_{1},\mathbf{p}_{2},\mathbf{K}) = \frac{\Omega}{N_{b}\sqrt{z}} P_{12} \sum_{\lambda_{1}\lambda_{2}\lambda_{3}} Z_{1\lambda_{1}}(\mathbf{p}_{1})Z_{1\lambda _{2}}(\mathbf{p}_{2})Z_{1\lambda_{3}}(\mathbf{p}_{3}) \frac{1}{2\left[\varepsilon_{\mathbf{p}_{3}\mathbf{+K}}-E_{1\lambda _{3}}(\mathbf{p}_{3})\right]} \nonumber \\
\fl \Bigg[ \frac{2T(\mathbf{q}-\mathbf{p}_{1},E_{3B}(\mathbf{q})-E_{1\lambda_{1}}(\mathbf{p}_{1}))f_{\mathbf{p}_{1}\lambda_{1}}+T(\mathbf{q}-\mathbf{p}_{3},E_{3B}(\mathbf{q})-\varepsilon_{\mathbf{p}_{3}\mathbf{+K}})F(\mathbf{p}_{3},\varepsilon_{\mathbf{p}_{3}\mathbf{+K}})}{E_{3B}(\mathbf{q})-E_{1\lambda_{1}}(\mathbf{p}_{1})-E_{1\lambda_{2}}(\mathbf{p}_{2})-\varepsilon_{\mathbf{p}_{3}\mathbf{+K}}} \nonumber \\
\fl -\frac{2T(\mathbf{q}-\mathbf{p}_{1},E_{3B}(\mathbf{q})-E_{1\lambda_{1}}(\mathbf{p}_{1})) f_{\mathbf{p}_{1}\lambda_{1}} + T(\mathbf{q}-\mathbf{p}_{3},E_{3B}(\mathbf{q})-E_{1\lambda_{3}}(\mathbf{p}_{3})) f_{\mathbf{p}_{3}\lambda_{3}}}{E_{3B}(\mathbf{q})-E_{1\lambda_{1}}(\mathbf{p}_{1})-E_{1\lambda_{2}}(\mathbf{p}_{2})-E_{1\lambda_{3}}(\mathbf{p}_{3})} \Bigg],\nonumber \\
\fl f_{baa}(\mathbf{p}_{1},\mathbf{p}_{2},\mathbf{K}_{2},\mathbf{K}_{3}) = \frac{\Omega^{2}}{N_{b}z} P_{23}\sum_{\lambda_{1}\lambda_{2}\lambda_{3}} \frac{Z_{1\lambda_{1}}(\mathbf{p}_{1})Z_{1\lambda_{2}}(\mathbf{p}_{2})Z_{1\lambda_{3}}(\mathbf{p}_{3})}{2(\varepsilon_{\mathbf{p}_{2}\mathbf{+K}_{2}}-E_{1\lambda _{2}}(\mathbf{p}_{2}))} \nonumber \\
\fl \Bigg[ \frac{T(\mathbf{q}-\mathbf{p}_{1},E_{3B}(\mathbf{q})-E_{1\lambda_{1}}(\mathbf{p}_{1}))f_{\mathbf{p}_{1}\lambda_{1}} + 2T(\mathbf{q}-\mathbf{p}_{2},E_{3B}(\mathbf{q})-\varepsilon_{\mathbf{p}_{2}\mathbf{+K}_{2}}) F(\mathbf{p}_{2},\varepsilon_{\mathbf{p}_{2}\mathbf{+K}_{2}})}{(E_{3B}(\mathbf{q})-E_{1\lambda_{1}}(\mathbf{p}_{1})-\varepsilon_{\mathbf{p}_{2}\mathbf{+K}_{2}}-\varepsilon_{\mathbf{p}_{3}\mathbf{+K}_{3}})(E_{3B}(\mathbf{q})-E_{1\lambda_{1}}(\mathbf{p}_{1})-\varepsilon_{\mathbf{p}_{2}\mathbf{+K}_{2}}-E_{1\lambda _{3}}(\mathbf{p}_{3}))} \nonumber \\
\fl -\frac{T(\mathbf{q}-\mathbf{p}_{1},E_{3B}(\mathbf{q})-E_{1\lambda_{1}}(\mathbf{p}_{1}))f_{\mathbf{p}_{1}\lambda_{1}}+2T(\mathbf{q}-\mathbf{p}_{2},E_{3B}(\mathbf{q})-E_{1\lambda_{2}}(\mathbf{p}_{2}))f_{\mathbf{p}_{2}\lambda_{2}}}{(E_{3B}(\mathbf{q})-E_{1\lambda_{1}}(\mathbf{p}_{1})-E_{1\lambda_{2}}(\mathbf{p}_{2})-\varepsilon_{\mathbf{p}_{3}\mathbf{+K}_{3}})(E_{3B}(\mathbf{q})-E_{1\lambda _{1}}(\mathbf{p}_{1})-E_{1\lambda_{2}}(\mathbf{p}_{2})-E_{1\lambda_{3}}(\mathbf{p}_{3}))} \Bigg], \nonumber \\
\fl f_{a}(\mathbf{p}_{1},\mathbf{p}_{2},K_{1},K_{2},K_{3}) = \frac{\Omega^{3}}{6N_{b}z\sqrt{z}} P_{123} \sum_{\lambda_{1}\lambda_{2}\lambda_{3}}\frac{Z_{1\lambda_{1}}(\mathbf{p}_{1})Z_{1\lambda_{2}}(\mathbf{p}_{2})Z_{1\lambda_{3}}(\mathbf{p}_{3})}{\varepsilon_{\mathbf{p}_{1}\mathbf{+K}_{1}}-E_{1\lambda_{1}}(\mathbf{p}_{1})} \nonumber \\
\fl \Bigg[ \frac{2E_{3B}(\mathbf{q})-2\varepsilon_{\mathbf{p}_{1}\mathbf{+K}_{1}}-\varepsilon_{\mathbf{p}_{2}\mathbf{+K}_{2}}-\varepsilon_{\mathbf{p}_{3}\mathbf{+K}_{3}}-E_{1\lambda_{2}}(\mathbf{p}_{2})-E_{1\lambda_{3}}(\mathbf{p}_{3})}{(E_{3B}(\mathbf{q})-\varepsilon_{\mathbf{p}_{1}\mathbf{+K}_{1}}-\varepsilon_{\mathbf{p}_{2}\mathbf{+K}_{2}}-\varepsilon_{\mathbf{p}_{3}\mathbf{+K}_{3}})(E_{3B}(\mathbf{q})-\varepsilon_{\mathbf{p}_{1}\mathbf{+K}_{1}}-\varepsilon_{\mathbf{p}_{2}\mathbf{+K}_{2}}-E_{1\lambda _{3}}(\mathbf{p}_{3}))} \nonumber \\
\fl \times \frac{T(\mathbf{q}-\mathbf{p}_{1},E_{3B}(\mathbf{q})-\varepsilon_{\mathbf{p}_{1}\mathbf{+K}_{1}})F(\mathbf{p}_{1},\varepsilon_{\mathbf{p}_{1}\mathbf{+K}_{1}})}{(E_{3B}(\mathbf{q})-\varepsilon_{\mathbf{p}_{1}\mathbf{+K}_{1}}-E_{1\lambda_{2}}(\mathbf{p}_{2})-\varepsilon_{\mathbf{p}_{3}\mathbf{+K}_{3}})(E_{3B}(\mathbf{q})-\varepsilon_{\mathbf{p}_{1}\mathbf{+K}_{1}}-E_{1\lambda_{2}}(\mathbf{p}_{2})-E_{1\lambda_{3}}(\mathbf{p}_{3}))} \nonumber \\
\fl -\frac{2E_{3B}(\mathbf{q})-2E_{1\lambda _{1}}(\mathbf{p}_{1})-\varepsilon_{\mathbf{p}_{2}\mathbf{+K}_{2}}-\varepsilon_{\mathbf{p}_{3}\mathbf{+K}_{3}}-E_{1\lambda_{2}}(\mathbf{p}_{2})-E_{1\lambda_{3}}(\mathbf{p}_{3})}{(E_{3B}(\mathbf{q})-E_{1\lambda _{1}}(\mathbf{p}_{1})-\varepsilon_{\mathbf{p}_{2}\mathbf{+K}_{2}}-\varepsilon_{\mathbf{p}_{3}\mathbf{+K}_{3}})(E_{3B}(\mathbf{q})-E_{1\lambda _{1}}(\mathbf{p}_{1})-\varepsilon_{\mathbf{p}_{2}\mathbf{+K}_{2}}-E_{1\lambda_{3}}(\mathbf{p}_{3}))} \nonumber \\
\fl \times \frac{T(\mathbf{q}-\mathbf{p}_{1},E_{3B}(\mathbf{q})-E_{1\lambda _{1}}(\mathbf{p}_{1}))f_{\mathbf{p}_{1}\lambda_{1}}}{(E_{3B}(\mathbf{q})-E_{1\lambda_{1}}(\mathbf{p}_{1})-E_{1\lambda_{2}}(\mathbf{p}_{2})-\varepsilon_{\mathbf{p}_{3}\mathbf{+K}_{3}})(E_{3B}(\mathbf{q})-E_{1\lambda _{1}}(\mathbf{p}_{1})-E_{1\lambda_{2}}(\mathbf{p}_{2})-E_{1\lambda_{3}}(\mathbf{p}_{3}))} \Bigg]\nonumber,
\end{eqnarray}
where the total momentum $\mathbf{q=p}_{1}+\mathbf{p}_{2}+\mathbf{p}_{3}$. The Fourier transforms of these wavefunctions give rise to the wavefunctions in the coordinate space shown in Sec.~\ref{sec:manyexcitation}.

\newpage

\section*{References}

\bibliographystyle{unsrt}
\bibliography{Sci,books}

\begin{thebibliography}{10}

\bibitem{cohenbook92a}
Claude Cohen-Tannoudji, Jacques Dupont-Roc, Gilbert Grynberg, and Patricia
  Thickstun.
\newblock {\em Atom-photon interactions: basic processes and applications}.
\newblock Wiley Online Library, 1992.

\bibitem{lehmberg70a}
R.~H. Lehmberg.
\newblock {Radiation from an $N$-Atom System. I. General Formalism}.
\newblock {\em Phys. Rev. A}, 2(3):883--888, Sep 1970.

\bibitem{lehmberg70b}
R.~H. Lehmberg.
\newblock Radiation from an $n$-atom system. ii. spontaneous emission from a
  pair of atoms.
\newblock {\em Phys. Rev. A}, 2:889--896, Sep 1970.

\bibitem{raimond01a}
J.~M. Raimond, M.~Brune, and S.~Haroche.
\newblock {Manipulating quantum entanglement with atoms and photons in a
  cavity}.
\newblock {\em {Rev. Mod. Phys.}}, 73:565, 2001.

\bibitem{ritsch13a}
Helmut Ritsch, Peter Domokos, Ferdinand Brennecke, and Tilman Esslinger.
\newblock Cold atoms in cavity-generated dynamical optical potentials.
\newblock {\em Rev. Mod. Phys.}, 85:553--601, Apr 2013.

\bibitem{bykov75a}
Vladimir~P Bykov.
\newblock Spontaneous emission from a medium with a band spectrum.
\newblock {\em Soviet Journal of Quantum Electronics}, 4(7):861, 1975.

\bibitem{john90a}
Sajeev John and Jian Wang.
\newblock {Quantum electrodynamics near a photonic band gap: Photon bound
  states and dressed atoms}.
\newblock {\em Phys. Rev. Lett.}, 64:2418--2421, May 1990.

\bibitem{kurizki90a}
Gershon Kurizki.
\newblock Two-atom resonant radiative coupling in photonic band structures.
\newblock {\em Phys. Rev. A}, 42:2915--2924, Sep 1990.

\bibitem{thompson13a}
J.~D. Thompson, T.~G. Tiecke, N.~P. de~Leon, J.~Feist, A.~V. Akimov,
  M.~Gullans, A.~S. Zibrov, V.~Vuletic, and M.~D. Lukin.
\newblock Coupling a single trapped atom to a nanoscale optical cavity.
\newblock {\em Science}, 340(6137):1202--1205, 2013.

\bibitem{goban13a}
A.~Goban, C.-L. Hung, S.-P Yu, J.D. Hood, J.A. Muniz, J.H. Lee, M.J. Martin,
  A.C. McClung, K.S. Choi, D.E. Chang, O.~Painter, and H.J. Kimblemblrm.
\newblock Atom-light interactions in photonic crystals.
\newblock {\em Nat. Commun.}, 5:3808, 2014.

\bibitem{lodahl15a}
Peter Lodahl, Sahand Mahmoodian, and S\o{}ren Stobbe.
\newblock Interfacing single photons and single quantum dots with photonic
  nanostructures.
\newblock {\em Rev. Mod. Phys.}, 87:347--400, May 2015.

\bibitem{hood16a}
Jonathan~D Hood, Akihisa Goban, Ana Asenjo-Garcia, Mingwu Lu, Su-Peng Yu,
  Darrick~E Chang, and HJ~Kimble.
\newblock Atom--atom interactions around the band edge of a photonic crystal
  waveguide.
\newblock {\em Proceedings of the National Academy of Sciences},
  113(38):10507--10512, 2016.

\bibitem{liu17a}
Yanbing Liu and Andrew~A Houck.
\newblock Quantum electrodynamics near a photonic bandgap.
\newblock {\em Nature Physics}, 13(1):48--52, 2017.

\bibitem{mirhosseini18a}
Mohammad Mirhosseini, Eunjong Kim, Vinicius~S Ferreira, Mahmoud Kalaee, Alp
  Sipahigil, Andrew~J Keller, and Oskar Painter.
\newblock Superconducting metamaterials for waveguide quantum electrodynamics.
\newblock {\em arXiv:1802.01708}, 2018.

\bibitem{sundaresan18a}
Neereja~M Sundaresan, Rex Lundgren, Guanyu Zhu, Alexey~V Gorshkov, and Andrew~A
  Houck.
\newblock Interacting qubit-photon bound states with superconducting circuits.
\newblock {\em arXiv:1801.10167}, 2018.

\bibitem{devega08a}
In{\'e}s de~Vega, Diego Porras, and J.~{Ignacio Cirac}.
\newblock {Matter-Wave Emission in Optical Lattices: Single Particle and
  Collective Effects}.
\newblock {\em Phys. Rev. Lett.}, 101:260404, Dec 2008.

\bibitem{navarretebenlloch11a}
Carlos Navarrete-Benlloch, Ines de~Vega, Diego Porras, and J~Ignacio Cirac.
\newblock {Simulating quantum-optical phenomena with cold atoms in optical
  lattices}.
\newblock {\em New Journal of Physics}, 13(2):023024, 2011.

\bibitem{krinner17a}
Ludwig Krinner, Michael Stewart, Arturo Pazmino, Joonhyuk Kwon, and Dominik
  Schneble.
\newblock Spontaneous emission in a matter-wave open quantum system.
\newblock {\em arXiv:1712.07791}, 2017.

\bibitem{agarwal77a}
G.~S. Agarwal, A.~C. Brown, L.~M. Narducci, and G.~Vetri.
\newblock Collective atomic effects in resonance fluorescence.
\newblock {\em Physical Review A}, 15(4):1613, 1977.

\bibitem{shahmoon13a}
Ephraim Shahmoon and Gershon Kurizki.
\newblock Nonradiative interaction and entanglement between distant atoms.
\newblock {\em Phys. Rev. A}, 87:033831, 2013.

\bibitem{douglas15a}
James~S Douglas, H~Habibian, C-L Hung, AV~Gorshkov, H~Jeff Kimble, and
  Darrick~E Chang.
\newblock Quantum many-body models with cold atoms coupled to photonic
  crystals.
\newblock {\em Nature Photonics}, 9(5):326--331, 2015.

\bibitem{gonzaleztudela15c}
Alejandro Gonz{\'a}lez-Tudela, C-L Hung, Darrick~E Chang, J~Ignacio Cirac, and
  HJ~Kimble.
\newblock Subwavelength vacuum lattices and atom--atom interactions in
  two-dimensional photonic crystals.
\newblock {\em Nature Photonics}, 9(5):320--325, 2015.

\bibitem{shi09a}
T~Shi and CP~Sun.
\newblock Two-photon scattering in one dimension by localized two-level system.
\newblock {\em arXiv:0907.2776}, 2009.

\bibitem{sanchezburillo14a}
E.~Sanchez-Burillo, D.~Zueco, J.~J. Garcia-Ripoll, and L.~Martin-Moreno.
\newblock Scattering in the ultrastrong regime: Nonlinear optics with one
  photon.
\newblock {\em Phys. Rev. Lett.}, 113:263604, Dec 2014.

\bibitem{shi16a}
Tao Shi, Ying-Hai Wu, A.~Gonz\'alez-Tudela, and J.~I. Cirac.
\newblock Bound states in boson impurity models.
\newblock {\em Phys. Rev. X}, 6:021027, May 2016.

\bibitem{calajo16a}
Giuseppe Calaj{\'o}, Francesco Ciccarello, Darrick Chang, and Peter Rabl.
\newblock Atom-field dressed states in slow-light waveguide qed.
\newblock {\em Physical Review A}, 93(3):033833, 2016.

\bibitem{pethickbook02a}
Christopher~J Pethick and Henrik Smith.
\newblock {\em Bose-Einstein condensation in dilute gases}.
\newblock Cambridge University Press, 2002.

\bibitem{white92a}
Steven~R. White.
\newblock Density matrix formulation for quantum renormalization groups.
\newblock {\em Phys. Rev. Lett.}, 69:2863--2866, Nov 1992.

\bibitem{schollwock11a}
Ulrich Schollw{\"o}ck.
\newblock The density-matrix renormalization group in the age of matrix product
  states.
\newblock {\em Ann. Phys.}, 326(1):96 -- 192, 2011.

\bibitem{haldane81a}
F.~D.~M. Haldane.
\newblock Effective harmonic-fluid approach to low-energy properties of
  one-dimensional quantum fluids.
\newblock {\em Phys. Rev. Lett.}, 47:1840--1843, Dec 1981.

\bibitem{calabrese04a}
Pasquale Calabrese and John Cardy.
\newblock Entanglement entropy and quantum field theory.
\newblock {\em Journal of Statistical Mechanics: Theory and Experiment},
  2004(06):P06002, 2004.

\bibitem{gonzaleztudela17a}
A.~Gonz\'alez-Tudela and J.~I. Cirac.
\newblock Directional emission of quantum emitter in two-dimensional structured
  reservoirs.
\newblock {\em Main paper}.

\bibitem{gonzaleztudela18c}
A.~Gonz\'alez-Tudela and J.~I. Cirac.
\newblock Exotic quantum dynamics and purely long-range coherent interactions
  in dirac conelike baths.
\newblock {\em Phys. Rev. A}, 97:043831, Apr 2018.

\end{thebibliography}

\end{document}